\documentclass[aps,12pt,showpacs,tightenlines,amssymb]{revtex4}
\usepackage{epsfig} \usepackage{bm}
\newcommand{\be}{\begin{equation}}\newcommand{\ee}{\end{equation}}
\newcommand{\bea}{\begin{eqnarray}}\newcommand{\eea}{\end{eqnarray}}

\def\littlespace{$\,$}
\def\unit#1{\hbox{\littlespace #1}} % use WITHOUT space after number

\def\halftext{\textstyle{\frac12}} \def\htx{{\halftext}}
\def\quartertext{{\textstyle{\frac14}}}
\def\oms{\omega_s}\def\omone{\omega_1}
\def\Nmax{N_{\hbox{\tiny max}}}
\def\uupdown#1#2{u^{(#1)}_{#2}}

\def\local{|\hbox{local}\rangle}
\def\localdag{\langle\hbox{local}|}
\def\parentinybox#1{{(\tinybox{#1})}}
\def\Qchain#1{Q_{#1}^\parentinybox{chain}}
\def\Qring#1{Q_{#1}^\parentinybox{ring}}
\def\Omchain#1{\Omega_{#1}^\parentinybox{chain}}
\def\Omring#1{\Omega_{#1}^\parentinybox{ring}}
\def\Seff{S_{\hbox{\tiny eff}}}
\def\Eqref#1{Eq.~(\ref{#1})}
\def\Eqrefs#1#2{Eqs.~(\ref{#1}) and~(\ref{#2})}
\def\tinybox#1{{\hbox{\tiny #1}}}
\def\nothing#1{#1}
\def\nothing#1{}
\def\omeff{\omega_{\tinybox{eff}}}
\def\GamGR{\Gamma_{\hbox{\tiny GR}}}

\def\tauZ{\tau_{_Z}}
\def\parenref#1{(\ref{#1})}
\def\ltsim{\hbox{\kern.25em\raise.5ex\hbox{$<$}\kern-.75em\lower.5ex
   \hbox{$\sim$}\kern.25em}}
\def\gtsim{\hbox{\kern.25em\raise.5ex\hbox{$>$}\kern-.75em\lower.5ex
   \hbox{$\sim$}\kern.25em}}

\newcommand{\labeld}[1]{ }

\begin{document}
%%%%%%%%%%%%%%%%%%%%%%%%%%%%%%%%%%%%%%%%%%%%%%%%%%%%%%%%%%%%%
\title{Structure and time-dependence of quantum breathers}
\author{L. S. Schulman,\footnote{Email: schulman@clarkson.edu}}
\affiliation{Physics Department, Clarkson University, Potsdam, New York 13699-5820, USA}
\author{D. Tolkunov,\footnote{Email: tolkunov@clarkson.edu}}
\affiliation{Physics Department, Clarkson University, Potsdam, New York 13699-5820, USA}
\author{E. Mihokova,\footnote{Email: mihokova@fzu.cz}}
\affiliation{Institute of Physics, Academy of Sciences of the Czech Republic, Cukrovarnick\'a~10, 162~53~Prague~6, Czech~Republic}
\date{\today}
%%%%%%%%%%%%%%%%%%%%%%%%%%%%%%%%%%%%%%%%%%%%%%%%%%%%%%%%%%%%%%%%%%%%%%%%
\begin{abstract}
Quantum states of a discrete breather are studied in two ways. One method involves numerical diagonalization of the Hamiltonian, the other uses the path integral to examine correlations in the eigenstates. In both cases only the central nonlinearity is retained. To reduce truncation effects in the numerical diagonalization, a basis is used that involves a quadratic local mode. A similar device is used in the path integral method for deducing localization. Both approaches lead to the conclusion that aside from quantum tunneling the quantized discrete breather is stable. 
\end{abstract}
%%%%%%%%%%%%%%%%%%%%%%%%%%%%%%%%%%%%%%%%%%%%%%%%%%%%%%%%%%%%%%%%%%%%%%%%
\pacs{31.15.Gy, 31.70.Hq, 05.45.-a, 63.20.Pw.}
\maketitle

%%%%%%%%%%%%%%%%%%%%%%%%%%%%%%%%%%%%%%%%%%%%%%%%%%%%%%%%%%%%%%%%%%%%%%%%
\section{Introduction\label{sec:intro}\labeld{sec:intro}}
%%%%%%%%%%%%%%%%%%%%%%%%%%%%%%%%%%%%%%%%%%%%%%%%%%%%%%%%%%%%%%%%%%%%%%%%
Nonlinearity creates localized structures through a variety of mechanisms. In a discrete lattice, extreme displacement of one or a few atoms pushes them to high oscillatory frequencies outside the range of ordinary (linear) phonons, so that the energy sequestered in these large excitations does not spread throughout the lattice. Such an excitation, known as a \emph{discrete breather} (DB) or \emph{intrinsic localized mode} (ILM), is a kind of soliton, although different in many ways from the moving, continuum soliton of Russell \cite{russellall}. The discrete breathers of primary interest to us, because of their relation to decay anomalies in doped alkali halides \cite{confine}, are not ``intrinsic'' in the sense that they are induced by excitation of the impurity, and translational invariance has been lost. For convenience, and to minimize acronyms, we refer to the excitations we study as ``breathers.''

There is an extensive literature on these breathers and an introduction can be found in \cite{campbell}. Because of the nonlinearity, a good deal of the theoretical work in this field is numerical. For systems well-approximated classically this is not a problem, but treating the system quantum mechanically demands potentially serious approximations.

An issue of experimental relevance is the lifetime of the quantum breather against decay. At the classical level, both analytical and numerical experience support an infinite lifetime, at least for the one-frequency breather \cite{flachwillis}. In a diatomic lattice there are two breather frequencies (in the phonon gap and above the optical phonons) and although there are indications that combinations of the modes can allow energy to leave via the phonons, in our simulations this does not take place in any perceptible way.

But classical stability says little about its quantum counterpart. Quantum tunneling bypasses classical constraints, and for a translationally invariant system, guarantees that bound states that are not of finite support become bands. Indeed such bands---as well as tunneling---have been studied for breathers in a homogeneous lattice \cite{wang, fleurovschillingflach, flachfleurovtunnel}. Other breather quantization approaches have also been taken, including semiclassical and field-theoretical \cite{breatherquant, takeno}. In the semiclassical approach one uses Einstein-Brillouin-Keller quantization on the KAM torus-confined classical trajectory. Naturally this yields a stable state, since, as for a 1-D state hidden behind a barrier, the classical trajectory does not sense the possibility of escape. Nevertheless, in principle it is entirely plausible that corrections to semiclassical behavior, which generally begin at order $\hbar^2$, would include the possibility of decay (although in one-dimension the tunneling decay correction is the much smaller $\exp(-[\hbox{positive constant}]/\hbar)$).

The focus of the present article is the stability of the quantized breather. At times it seems that the existence of even the classical breather is magical. Were this not a known phenomenon, one's first reaction on finding this numerically should be to debug the program! Why then should one expect this ``magic'' to operate in the quantum case as well? Perhaps to first order in $\hbar$, where semiclassical considerations are a guide, but a priori there is no reason that quantum corrections should not lead to decay channels.

Indeed, in a series of articles \cite{hz1, hz2, hz3, hz4} it has been claimed that such decay takes place, and estimates for its rate in alkali halides give lifetimes of about 10\unit ns. Decay is found to occur irrespective of dimension and, as far as we can tell, the decay rate is of order $\hbar$, surprisingly large. We recall the framework used in these articles, since to some extent we will take a similar approach. It is typical of the classical breather that a small number of atoms (often just one) vibrate strongly while the others hardly move at all. This strong vibration should not be treated by perturbation theory, and one replaces this dynamic, nonlinear object by something more tractable. Refs.\ \cite{hz1, hz2, hz3, hz4} replace it by a fixed, rapidly oscillating forcing term which couples---including nonlinear coupling---to the phonon field of the crystal. Two approximations are inherent in this: nonlinear interactions other than those with the central atom are neglected and the motion of the central atom is treated classically, subject neither to quantum laws nor to the back reaction of the phonon field.

Recent work \cite{flachfleurovradiation} takes a similar perspective but comes to different conclusions regarding decay rates. For the infinite lattice the authors find no decay at all, with departures from this situation for finite $N$ (lattice size) going like $1/N$. In any case, this is in strong disagreement with nanosecond-scale lifetimes in physical crystals.

In the present article we treat the entire system quantum mechanically. We use two distinct approaches, having at their core fundamentally different approximations. In both cases however the unavoidably large impact of the breather is treated by introducing a \emph{local mode}, either for comparison or for use as a zeroth-order system around which to perturb. As in other studies we drop nonlinear contributions except those that involve the particles most active in the breather.

Our first method employs the Feynman path integral. An early triumph of that technique was the calculation of polaron properties \cite{feynmanpibook, feynmanstatmechbook}, and the principle lying behind its success was that almost all forces involved were linear. Using the path integral most degrees of freedom could be integrated, leaving the polaron with a self-coupling that was non-local in time and reflected the back-reaction of the lattice to its motion. All this was at the quantum level. By dropping all nonlinear terms except that of the most active breather atom we arrive at a situation where the same method can be used. This also works if the active atom has only linear interactions, and allows us to compare properties of the breather to properties of a local mode, one whose motion is known to be fully localized. This is important because it is not enough to show the existence of eigenstates that correspond to the breather---one must show these eigenstates to be localized. What emerges from our calculations is that correlations of the breather-atom's motion with other atoms in the lattice drop in the same way as those of a local mode, as the strength of the appropriate coupling constant is increased. The path integral method has the additional advantage that its extension to higher dimension should be possible. (Note that our use of the path integral is unrelated to certain other uses of functional integration for solitons, for example as described in the work of Faddeev and Korepin \cite{faddeev}.)

The second method is direct diagonalization of the Hamiltonian. Essentially the same calculation was performed in \cite{wang}, although they maintained translational invariance---they did not drop any nonlinear terms---and as a result obtained band structure. The difficulty of numerical diagonalization is that a cutoff on the size of the phonon Hilbert space is required. In a phonon basis, a quartic interaction couples to ever-higher phonon excitations, and perforce some of these must be neglected. Moreover, the cutoff must be fairly low. The elegance of the number-operator representation allows the analytic calculator to forget that the phonon spaces form a tensor product. Nevertheless, if $n$ phonons are considered, with cutoff Hilbert spaces of dimension $N_c$, the Hamiltonian will live in a space of ${N_c}^n$ dimensions, potentially a computational disaster. Our innovation is to introduce a local mode into the zeroth order Hamiltonian, one whose frequency is roughly that of the breather in which we are interested. In this way, when going over to nonlinear coupling very few phonons are excited other than those of the local mode, and a low cutoff does little harm.

Our conclusion from these calculations is that the quantum breather, at least in one dimension, is stable. (One-dimension, because of the intense distortion of the symmetry-breaking Jahn-Teller effect, is the case of physical interest to us \cite{confine}.) From the path integral, the matching of correlations shows that the quantum breather has the same localization properties as a local mode, which is known to have an excitation profile (for atoms of the lattice) that drops off exponentially away from its source (as will be explicitly shown). For the direct diagonalization, the contribution of phonons other than those of the local mode to the true eigenstate is extremely small, on the order of $10^{-3}$ for the parameters we use. We also address the issue of whether even very small contributions can lead to decay, which might seem to be implied by Fermi's Golden Rule. It turns out that coupling to a continuum does not always mean decay to the continuum, as was found in \cite{limited}. In the present article we give explicit examples and in particular examine the time evolution of a state initially in an eigenstate of the local mode---which is \emph{not} an eigenstate of the full Hamiltonian. It does not decay.

In Sec.\ \ref{sec:dropnonlinearity} we show that dropping nonlinear terms, except for those affecting the central atom, leaves the classical mechanics qualitatively unchanged. Following that, in Sec.\ \ref{sec:pathintegral}, we present the path integral approach. Direct diagonalization of the Hamiltonian appears in Sec.\ \ref{sec:diagonalization} and the associated time evolution in Sec.\ \ref{sec:timeevolution}. A final section is devoted to discussion and conclusions.

%%%%%%%%%%%%%%%%%%%%%%%%%%%%%%%%%%%%%%%%%%%%%%%%%%%%%%%%%%%%%%%%%%%%%%%%
\section{Definitions and notation\label{sec:definitions}\labeld{sec:definitions}}
%%%%%%%%%%%%%%%%%%%%%%%%%%%%%%%%%%%%%%%%%%%%%%%%%%%%%%%%%%%%%%%%%%%%%%%%
The system is a ring of $N+1$ unit-mass atoms with Hamiltonian
\be
H=\sum_{k=0}^N \left\{\htx p_k^2 +\htx\omega_s^2 x_k^2 
             +\htx \omega_0^2(x_k-x_{k+1})^2
                    +\quartertext\lambda x_k^4\right\}
\,,\qquad x\in{\mathbb R}\,,
\label{eq:H1all}
\ee
\labeld{eq:H1all} 
Periodicity is expressed through $x_0\equiv x_{N+1}$ and mod-($N$+1) addition for atom labels. This is the nonlinearity studied by \cite{wang}. Another system, closer to our own models, is
\be
H=\sum_{k=0}^N \left\{\htx p_k^2 +\htx\omega_s^2 x_k^2 
             +\htx \omega_0^2(x_k-x_{k+1})^2
                  +\quartertext\lambda (x_k-x_{k+1})^4\right\}
\,,
\label{eq:H2all}
\ee
\labeld{eq:H2all} also periodic. To avoid being awash in subscripts and superscripts we distinguish these by context, rather than by labels attached to $H$. As remarked, when a breather is present, one typically has a single atom where most of the energy is concentrated, with the other atoms relatively still. We take the dynamic atom to be \#0 and consider an alternative Hamiltonian in which all nonlinear terms that do not involve $x_0$ are dropped. The corresponding Hamiltonians are
\bea
H&=&\sum_{k=0}^N \left\{\htx p_k^2 +\htx\omega_s^2 x_k^2 
             +\htx \omega_0^2(x_k-x_{k+1})^2\right\}
             +\quartertext\lambda x_0^4               \,,
\label{eq:H1} \\
H&=&\sum_{k=0}^N \left\{\htx p_k^2 +\htx\omega_s^2 x_k^2 
             +\htx \omega_0^2(x_k-x_{k+1})^2\right\}
             +\quartertext\lambda\left[ (x_0-x_{1})^4+ (x_0-x_{N})^4\right]
\,.
\label{eq:H2}
\eea
\labeld{eq:H1}\labeld{eq:H2}
For each of these we introduce a \emph{local mode} Hamiltonian, namely a linear system which, at a level to be explored below, behaves similarly to those in \Eqrefs{eq:H1}{eq:H2}. They are
\bea
H_0&=&\sum_{k=0}^N \left\{\htx p_k^2 +\htx\omega_s^2 x_k^2 
             +\htx \omega_0^2(x_k-x_{k+1})^2\right\}
             +\htx\omega_1^2 x_0^2
\,,
\label{eq:H01}\\
H_0&=&\sum_{k=0}^N \left\{\htx p_k^2 +\htx\omega_s^2 x_k^2 
             +\htx \omega_0^2(x_k-x_{k+1})^2\right\}
             +\htx\omega_1^2\left[ (x_0-x_{1})^2+ (x_0-x_{N})^2\right]
\,.
\label{eq:H02}
\eea
\labeld{eq:H01}\labeld{eq:H02}
As the zero subscript suggests, these will be the zero-order perturbation Hamiltonians in our numerical diagonalization. The perturbations will then have the form
\bea
V_I&=& \quartertext\lambda x_0^4 -\htx\omega_1^2 x_0^2
\,,
\label{eq:V1}\\
V_I&=& \quartertext\lambda\left[ (x_0-x_{1})^4+ (x_0-x_{N})^4\right]
        -\htx\omega_1^2\left[ (x_0-x_{1})^2+ (x_0-x_{N})^2\right]
\,,
\label{eq:V2}
\eea
\labeld{eq:V1}\labeld{eq:V2}
for the $H_0$'s of Eqs.\ \parenref{eq:H01} and \parenref{eq:H02}, respectively.

All these systems have reflection symmetry about atom-0. Define $P$ by
\be
P: k\leftrightarrow (N-k+1) \,,\ k=0,\ldots,\Nmax \hbox{~with~} \Nmax=\left[{\textstyle\frac {N+1}2}\right] \,,
\label{eq:P}
\ee
\labeld{eq:P}
and the square brackets represent ``integer part of.'' Classically an initial condition having this symmetry will retain it. Quantum mechanically, since $V_I$ shares this symmetry, only phonons even under $P$ need be considered.

%%%%%%%%%%%%%%%%%%%%%%%%%%%%%%%%%%%%%%%%%%%%%%%%%%%%%%%%%%%%%%%%%%%%%%%%
\section{Effect of dropping nonlinearity for ``non-breather'' atoms\label{sec:dropnonlinearity}\labeld{sec:dropnonlinearity}}
%%%%%%%%%%%%%%%%%%%%%%%%%%%%%%%%%%%%%%%%%%%%%%%%%%%%%%%%%%%%%%%%%%%%%%%%

\begin{figure}
\includegraphics[height=.4\textheight,width=.7\textwidth]{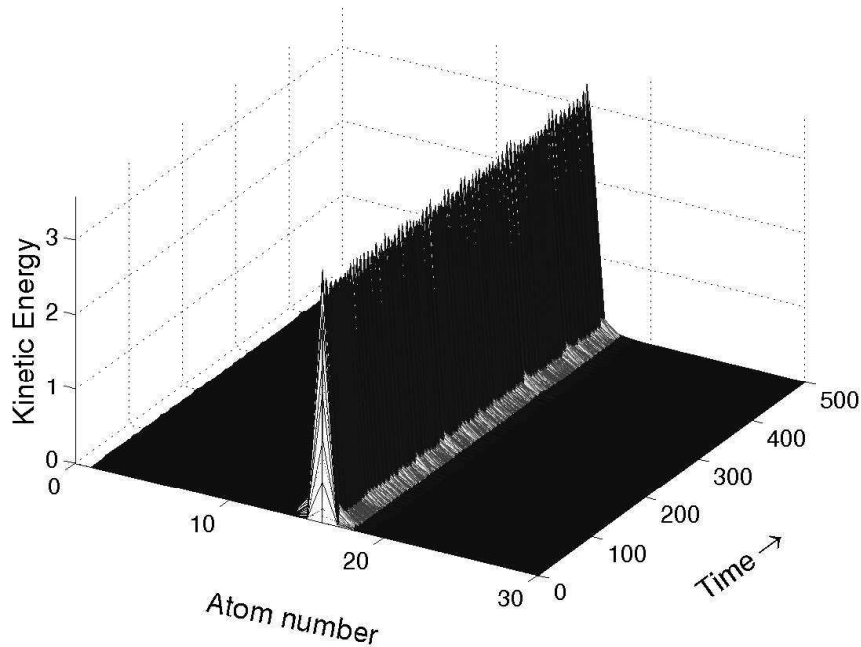}
\caption{Kinetic energy as a function of time. \textit{Only one figure is shown, as the others are indistinguishable at this level of display.}}
\label{fig:KE}\labeld{fig:KE}
\end{figure}

Since we drop nonlinear terms that do not include $x_0$, we first study the impact of that change on the classical mechanics. The dynamical systems to be compared are
\begin{itemize}
\item $H$ of \Eqref{eq:H1all}, ``All nonlinear.''
\item $H$ of \Eqref{eq:H1}, ``One nonlinear'' ($x_0$ in the equation).
\item $H_0$ of \Eqref{eq:H01}, ``Local mode'' (around $x_0$). 
\end{itemize}

We also make similar comparisons for nonlinear \textit{coupling}:
\begin{itemize}
\item $H$ of \Eqref{eq:H2all}, ``All nonlinear.''
\item $H$ of \Eqref{eq:H2}, ``One nonlinear'' ($x_0$ couples nonlinearly to its neighbors).
\end{itemize}

\begin{figure}\centerline{
\includegraphics[height=.2\textheight,width=.3\textwidth]{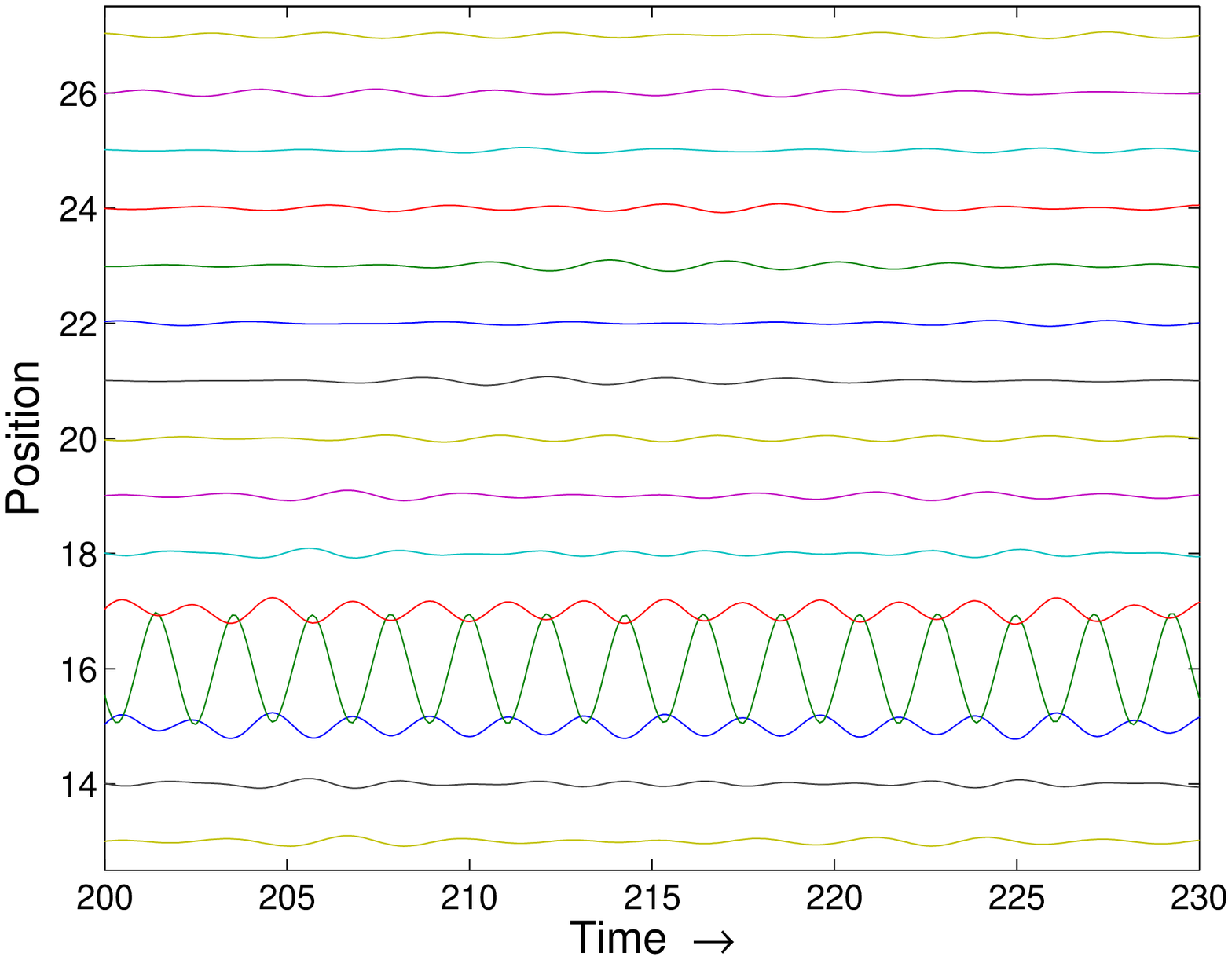}
\includegraphics[height=.2\textheight,width=.3\textwidth]{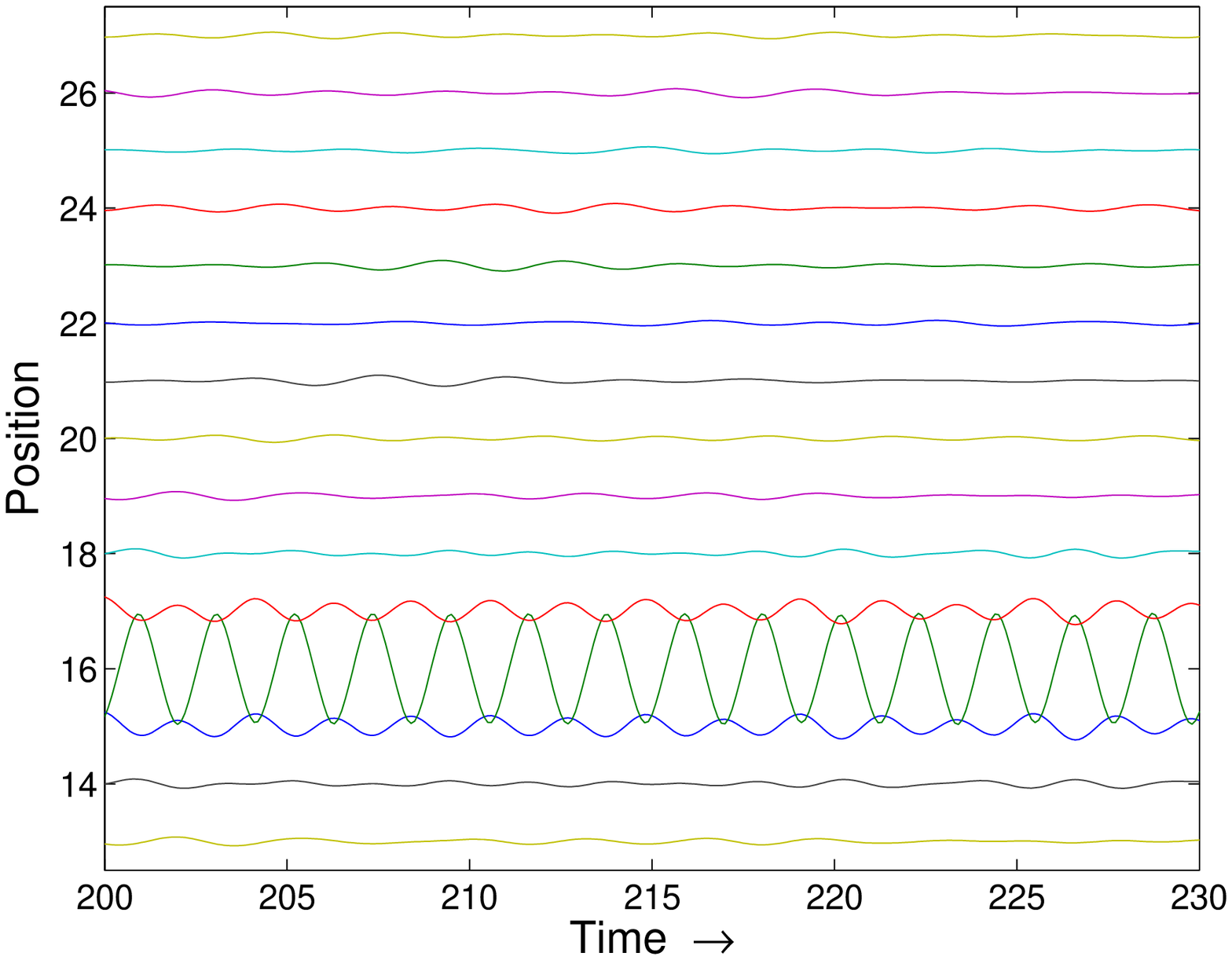}
\includegraphics[height=.2\textheight,width=.3\textwidth]{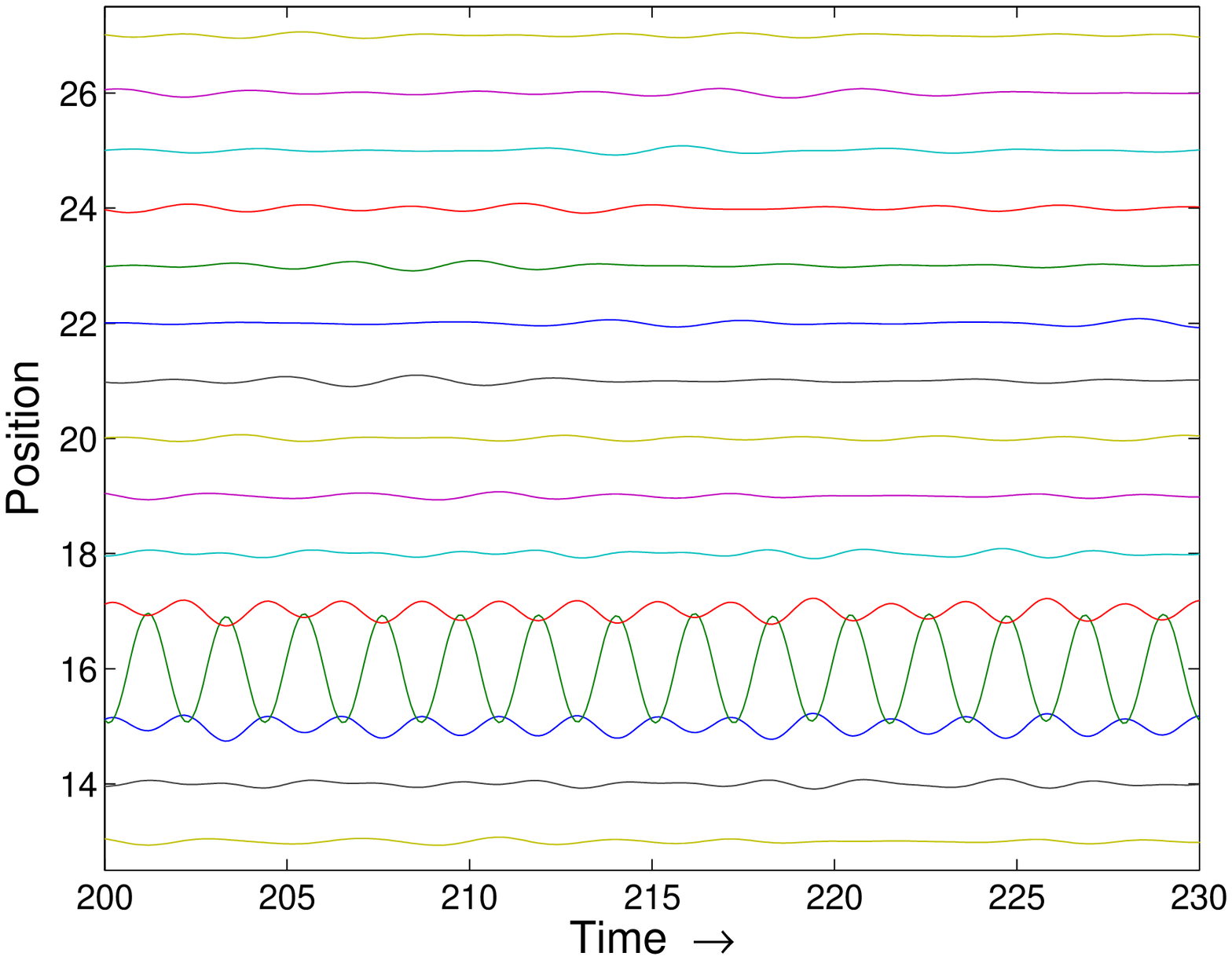}}
\caption{Position as a function of time. The motion of each atom is displaced by one, with the active breather atom in position~16. The figures illustrate symmetry and rapid dropoff as well similarity among the various dynamical systems. a) Full nonlinear system. b) Only nonlinearity in breather atom. c) Local mode at breather atom.}
\label{fig:positiondetail}\labeld{fig:positiondetail}
\end{figure}

\begin{figure}
\includegraphics[height=.25\textheight,width=.5\textwidth]{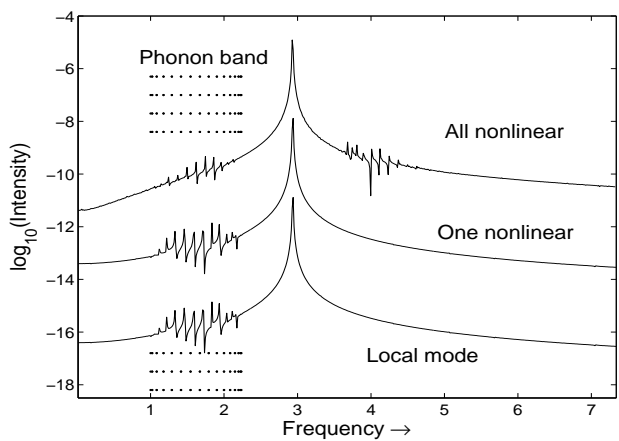}
\caption{Fourier transform of the position of first atom. The ``One nonlinear'' plot is displaced downward by 3 and the ``Local mode'' plot by 6. Dotted lines indicate the span of the phonon band [which does not begin at 0 because of the $\omega_s$ of Eqs. \parenref{eq:H1all} through \parenref{eq:H02}].}
\label{fig:spectfirstatom}\labeld{fig:spectfirstatom}
\end{figure}

For the self-coupling cases, we first exhibit the kinetic energy as a function of time for a ring of 30 atoms. As for our subsequent graphs we have shifted the central atom to the middle of the ring to avoid cutting the breather. The kinetic energy is a clear way to see the breather and is shown in Fig.\ \ref{fig:KE}. In Fig.\ \ref{fig:positiondetail}, we present position as a function of time for all three cases. Finally to illustrate the fact that differences do remain, we show the Fourier transform of the position of the central breather particle. Fig.\ \ref{fig:spectfirstatom} shows intensity, and it is clear that when \textit{all} atoms see the nonlinear force there are more pronounced higher harmonics in the motion of the central atom as well. All show activity in the phonon band, notwithstanding the fact that all are stable. 

\begin{figure}
\includegraphics[height=.4\textheight,width=.7\textwidth]{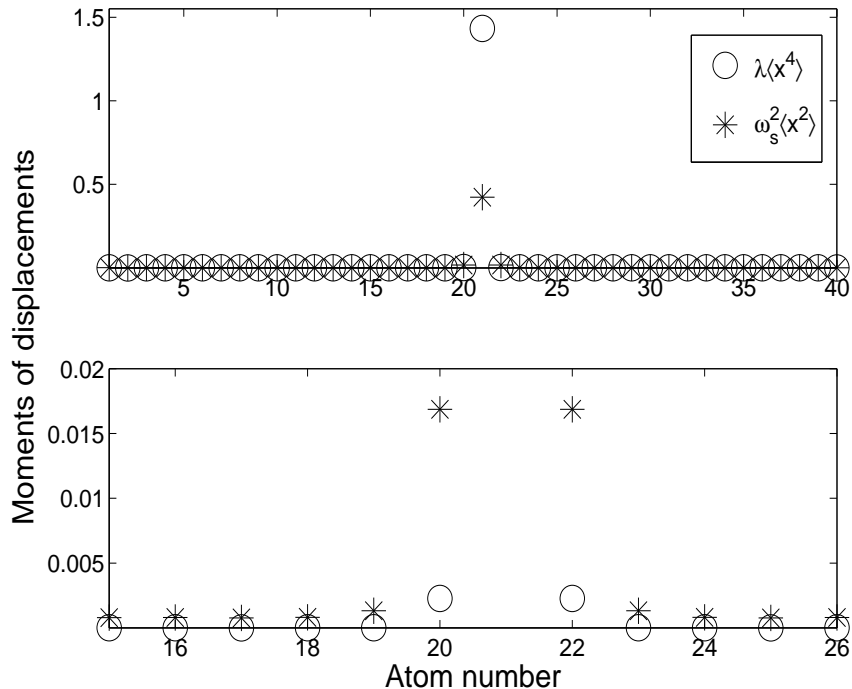}
\caption{Comparison of linear and nonlinear contributions to the energy. Nonlinearity is retained for \textit{all} atoms. The upper figure shows the time averages of $\omega_s^2x^2$ and $\lambda x^4$ for all atoms. The lower figure shows a subset of the same information, but on a scale large enough to see that even for the nearest neighbors of the most active atom the nonlinear contribution is extremely small.}
\label{fig:meandisplacement}\labeld{fig:meandisplacement}
\end{figure}

The reason for the relative indifference of breather behavior to nonlinear couplings not involving \#0 is that in any case atoms other than \#0 have small amplitude and hardly feel the nonlinearity. It is certainly true that for smaller $\lambda$, when the breather spreads over a large number of atoms (as occurs in several cases in \cite{bigconfine}), this approximation would be inadequate. However, in this article we consider only the highly localized case so as to focus on the issue of in-principle quantum stability. Fig.\ \ref{fig:meandisplacement}, based on the full Hamiltonian \Eqref{eq:H1all}, shows how small the nonlinear contributions to the energy are for any but the central breather atom. What are displayed are the time averages of $\omega_s^2x^2$ and $\lambda x^4$. It is clear that even for the nearest neighbors of the highly oscillatory atom the nonlinear contribution to the energy is negligible.

\begin{figure}\centerline{
\includegraphics[height=.25\textheight,width=.35\textwidth]{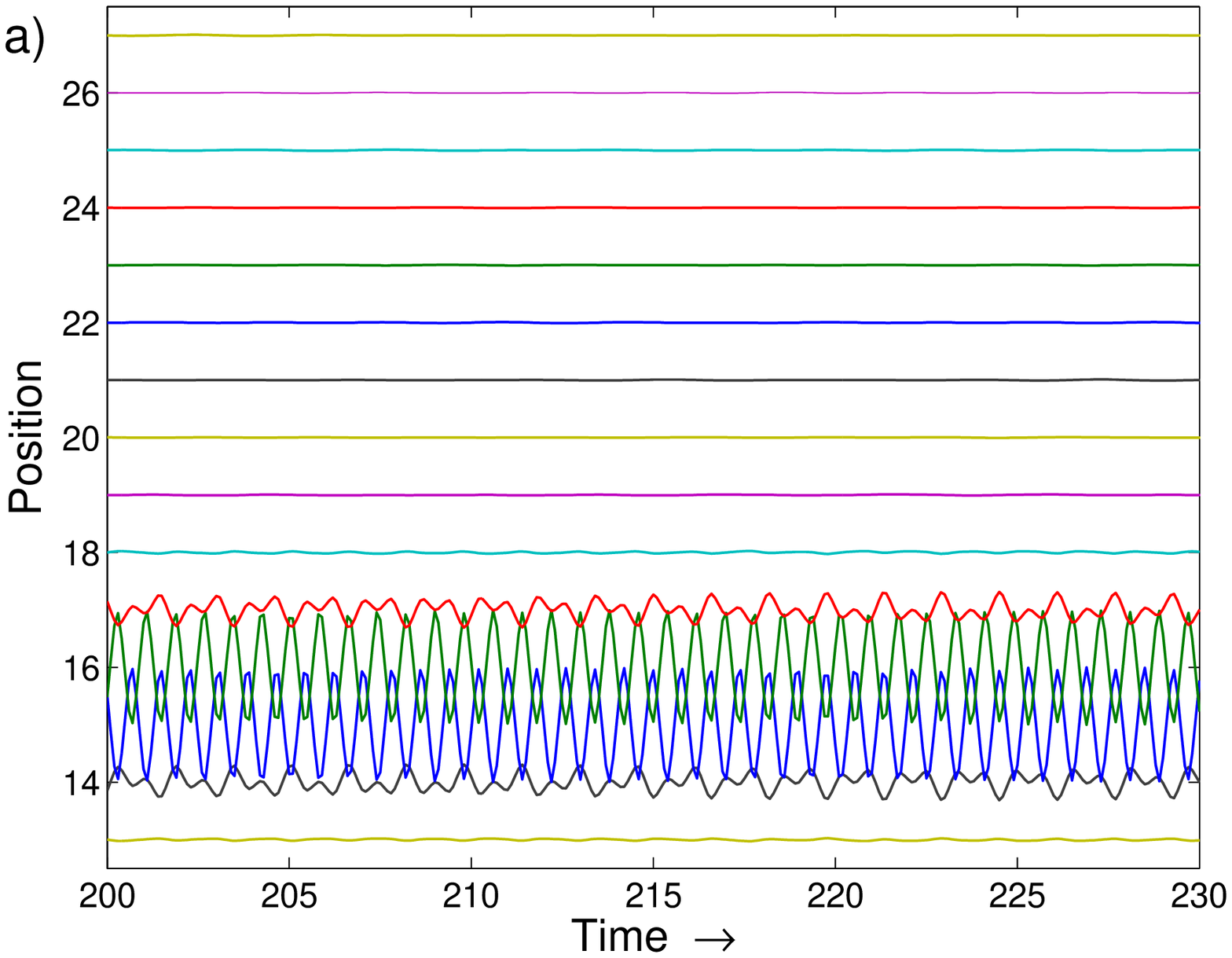}
\includegraphics[height=.25\textheight,width=.35\textwidth]{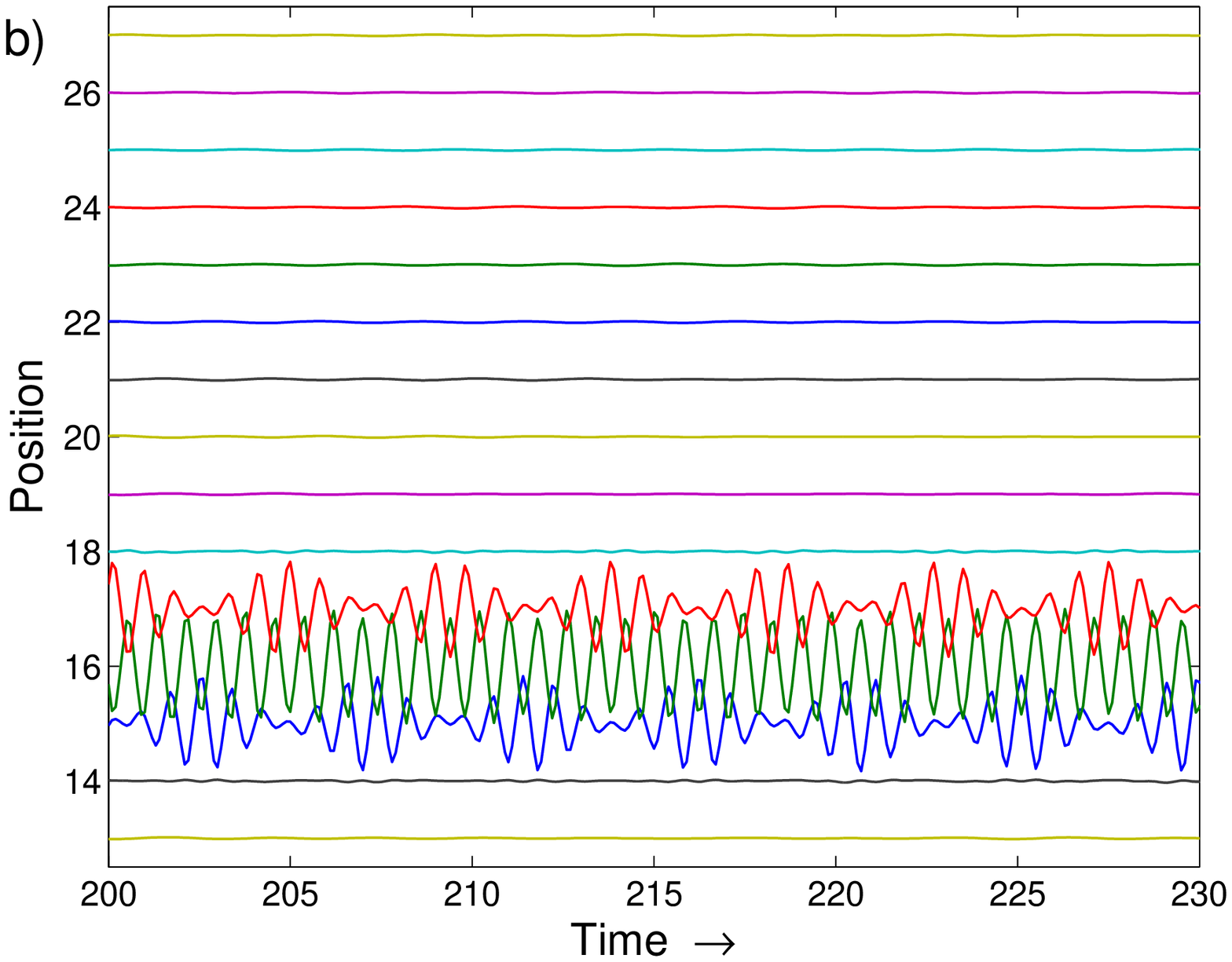}}
\caption{As in Fig.\ \ref{fig:positiondetail}. The figures illustrate symmetry and rapid dropoff as well as the differences between the full and truncated Hamiltonian. a) Full nonlinear system. b) Only nonlinearity in breather atom coupling.}
\label{fig:positiondetailcoupled}\labeld{fig:positiondetailcoupled}
\end{figure}

\begin{figure}
\includegraphics[height=.25\textheight,width=.35\textwidth]{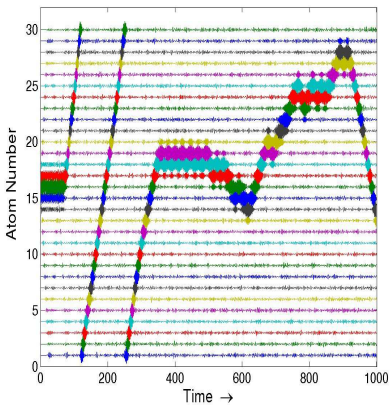}
\caption{Positions as a function of time for a ``wanderer.'' The motion is governed by the same equations that yield Fig.\ \ref{fig:positiondetailcoupled}, but the initial conditions are different. Here \emph{one} atom is displaced by a relatively large amount. }
\label{fig:wanderer}\labeld{fig:wanderer}
\end{figure}

For the case of the truncation of nonlinear \textit{coupling}, the parallel is not so close. The Hamiltonian, \Eqref{eq:H2all}, gives rise to interesting patterns of behavior not seen in the self-coupling case. 
%%%%%%%%%%%%%%%%%%%%%%%%%%%%%%%%%%%
There is first the phenomenon of moving breathers \cite{takenohori, horitakeno, burlakov, chubykalo, flachprl72, marin}, which obviously will not occur when only one site and its neighbors are subject to nonlinearity. Movement can be suppressed by taking appropriate initial conditions (equal and opposite position displacements) so that the stationary breather consists of a pair of atoms oscillating with respect to each other. To reproduce this in a truncated Hamiltonian we would have to keep the nonlinearity for at least two atoms, which, for our later numerical diagonalization, puts a strain on computer memory. We therefore stayed with the form \parenref{eq:H2}, which provides a local, nonlinear structure. The logic of \cite{hz1, hz2, hz3, hz4} predicts decay of this system when quantized, so that finding the eigenvectors of this system to be localized provides a further counterexample to that logic. In Fig.\ \ref{fig:positiondetailcoupled} we show the contrasting behavior of position as a function of time for the true breather and the truncated one. Apropos moving breathers, in Fig.\ \ref{fig:wanderer} we present an illustration of a typical example, wandering around the ring in a seemingly random way.

%%%%%%%%%%%%%%%%%%%%%%%%%%%%%%%%%%%%%%%%%%%%%%%%%%%%%%%%%%%%%
\subsection{Quantum perspective\label{sec:quantumperspective}\labeld{sec:quantumperspective}}
%%%%%%%%%%%%%%%%%%%%%%%%%%%%%%%%%%%%%%%%%%%%%%%%%%%%%%%%%%%%%

Quantum mechanically, dropping nonlinearity away from the center of the breather has a profound effect. It represents a loss of translational invariance, and, because of quantum tunneling, a qualitative change in the spectrum of the system. It follows that, in principle at least, for a translationally invariant system all breathers are nonlocal. The associated band structure for breather levels was found in \cite{wang} and Ref.\ \cite{fleurovchaos} has an extensive discussion of the significance of translational invariance.

It is therefore important to state what we mean in asserting that the breather is localized, and why our assertion is physically significant. If one considers a deeply bound atomic level, say the innermost shell of Cu when the atom is part of a crystal, that level---in principle---is part of a band. Practically, however, the band is irrelevant and the measurable properties of this level are independent of the near-zero tunneling probabilities that induce the theoretical band structure. In our work, we will consider breathers under circumstances where they are far from classical instability, so that quantum processes that connect to distinct classical breather states (e.g., the breather translated by one atom) would be of extremely small amplitude \cite{note:singularphasespace}. So we are saying that \emph{except for this small tunneling amplitude} the breather is stable. Physically this means that the breather can have an extremely long lifetime, contrary to the claims made in the literature that we have cited \cite{hz1, hz2, hz3, hz4}. We emphasize that in that literature translational invariance is also dropped.

A second consideration is that the physical systems of particular interest to us \cite{confine} do \emph{not} have translational invariance. The breather formed by Jahn-Teller distortion at an impurity certainly has a preferred origin, and it is stability against low order phonon decay that concerns us.
%%%%%%%%%%%%%%%%%%%%%%%%%%%%%%%%%%%%%%%%%%%%%%%%%%%%%%%%%%%%%
\section{The Path integral approach\label{sec:pathintegral}\labeld{sec:pathintegral}}
%%%%%%%%%%%%%%%%%%%%%%%%%%%%%%%%%%%%%%%%%%%%%%%%%%%%%%%%%%%%%%%%

Feynman's use of the path integral for the polaron \cite{feynmanpolaron, feynmanpibook} exploited what has come to be one of the most useful features of the path integral: the propagator for a quadratic degree of freedom coupled linearly to something else can be evaluated explicitly, leaving only a self-coupling of that ``something else.'' The penalty is the nonlocality in time of the self coupling, the reward the reduction of the problem to a single degree of freedom.

%%%%%%%%%%%%%%%%%%%%%%%%%%%%%%%%%%%%%%%%%%%%%%%%%%%%%%%%%%%%%%%
\subsection{Setting up the path integral\label{sec:settinguppathintegral}\labeld{sec:settinguppathintegral}}
%%%%%%%%%%%%%%%%%%%%%%%%%%%%%%%%%%%%%%%%%%%%%%%%%%%%%%%%%%%%%%%

\begin{figure}[tbp]
\includegraphics[scale=0.5]{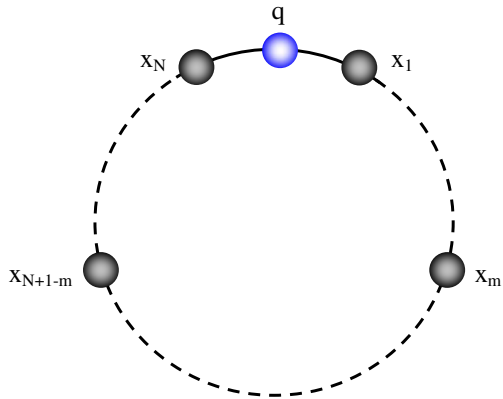}
\caption{Selected atoms on the ring. For convenience the notation $q$ is used for $x_0$.} \label{fig:ring} \labeld{fig:ring}
\end{figure}

As discussed in Sec.\ \ref{sec:dropnonlinearity}, for appropriate parameters, all atoms but one have negligible nonlinear energy contributions, and we drop the ``$\lambda x^4$'' for all but one of them. This allows us to integrate all but the single atom, around which the breather is centered. As in \Eqref{eq:H1}, we take that atom to be \#0. As a result we consider the Lagrangian
\be
{\cal L}
  =\frac12\sum\limits_{n=1}^{N+1}\dot x_n^2
     -\frac 12\omega_0^2\sum\limits_{n=1}^{N+1}(x_n-x_{n+1})^2
     -\frac 12\omega_s^2\sum\limits_{n=1}^{N+1}x_n^2-\frac \lambda 4 x_{0}^4  +\mu x_0(x_m+x_{N+1-m})\,.
\label{eq:lagrangian}
\ee
\labeld{eq:lagrangian}
We have introduced an additional term, $\mu x_0(x_m+x_{N+1-m})$. It is easy to carry along and will later allow us to study correlations. The derivative (at $\mu=0$) of the (appropriate form of the) propagator provides a (0-$m$)-correlation. We will take particle-$m$ to be distant from 0, typically about $1/3$ of the way around the ring. The use of a pair [$m$ and $(N+1-m)$] maintains mirror symmetry (``$P$'') with respect to particle-0. We will also distinguish between the \emph{ring}, all $N+1$ atoms, and the \emph{chain}, atoms 1 through~$N$. See Fig.\ \ref{fig:ring}.

The Lagrangian $\mathcal{L}$ can be written in the form $ \mathcal{L}=\mathcal{L}_0+\mathcal{L}_C+\mathcal{L}_I$. Here  $\mathcal{L}_0$ is the Lagrangian of particle-0
\be
\mathcal{L}_0=\frac 12\dot{q}^2-\left( \omega _0^2+\frac 12\omega_s^2\right) q^2-\frac \lambda 4q^4,
\label{eq:lagrangian0}
\ee
\labeld{eq:lagrangian0}
where we have adopted the notation $q$ for $x_0$. $\mathcal{L}_C$ is the Lagrangian of the chain
\be
\mathcal{L}_C
          =\frac12\sum\limits_{n=1}^N\dot{x}_n^2
          -\frac 12\omega_0^2\sum\limits_{m,n=1}^NJ_{mn}x_mx_n-\frac 12\omega_s^2\sum\limits_{n=1}^Nx_n^2 \,,
\label{eq:lagrangianR}
\ee
\labeld{eq:lagrangianR}
with $J_{mn}=2\delta _{mn}-\delta _{m+1,n}-\delta _{m,n+1}$ ($m,n=1\ldots N$), the Jacoby matrix. The last term $\mathcal{L}_I$ is the interaction between the particle-0 and the others
\be
\mathcal{L}_I
             =\omega _0^2q\left( x_1+x_N\right) +\mu q(x_m+x_{N+1-m}) \,.
\label{eq:lagrangianI}
\ee
\labeld{eq:lagrangianI}

We next perform the path integral for the chain and arrive at an effective action and path integral for $q$ alone \cite{feynmanpibook, feynmanstatmechbook, lsspibook, weiss}. Normal coordinates for the chain are defined by $\mathbf{Q}= \mathbf{Sx}$, where $S_{mn}=\sqrt{2/\left( N+1\right)}\sin \left[\pi mn/\left(N+1\right)\right]$ is a unitary matrix, and $(x_1,\dots,x_N)$ is considered a column vector. This leads to
\be
\mathcal{L}_C+\mathcal{L}_I
                        =\frac 12\sum\limits_{n=1}^N\left\{Q_n^2
                        -\Omega _n^2Q_n^2+f_n(t,q)Q_n\right\} \,.
\label{eq:lagrangianRI}
\ee
\labeld{eq:lagrangianRI}
Here $\{\Omega_n\}$ are the \emph{chain} normal-mode frequencies
\be
\Omega _n^2=\omega_s^2+4\omega _0^2\sin ^2\frac{\pi n}{2\left( N+1\right)} \;,
\label{eq:OmegaChain}
\ee
\labeld{eq:OmegaChain}
and $f_n(t,q)=q\tau _n\left[ 1-\left(-1\right) ^n\right]$, where
\be
\tau _n\equiv \sqrt{\frac 2{N+1}}\left( \omega _0^2\sin \frac{\pi n}{N+1}
             +\mu \sin \frac{\pi nm}{N+1}\right) \,.
\label{eq:tau}
\ee
\labeld{eq:tau}

The path integrals over the normal chain coordinates $Q_n$ can easily be evaluated, so that one has a propagator that is a function of the endpoints of \emph{all} degrees of freedom. It is convenient to take the matrix element of this object in the chain ground state and divide by corresponding matrix elements of the unforced chain oscillators. This yields an effective action for the chain
\be
e^{\frac i\hbar S_{\hbox{\tiny eff}}[q]}
       =\left[ \oint\prod_{n=1}^N\mathcal{D}Q_n\langle\psi _0^f|e^{\frac i\hbar (S_C+S_I)}|\psi_0^i\rangle\right]
       \Big/\oint \prod_{n=1}^N\mathcal{D}Q_n\langle\psi_0^f|e^{\frac i\hbar S_C}|\psi _0^i\rangle\,,
\label{eq:ExpSeff}
\ee
\labeld{eq:ExpSeff}
where $S_\gamma=\int{\cal L}_\gamma$, $\gamma=C,I$, and (e.g.) $\psi_0^f$ is shorthand for $\prod_n\psi_0^{(\Omega_n)}(Q_n^f)$, and $\psi_0^{(\omega)}(Q)=\left(\omega /\pi \hbar \right)^{1/4}\exp(-\omega Q^2/2\hbar)$ is the ground state of a harmonic oscillator. The chain oscillator modes are subject to a forcing term due to the so-far-unintegrated $q(t)$.

The result of these standard calculations is that the propagator for particle-0 takes the form
\be
\mathcal{G}\left(q^f,T;q^i,0\right)
          =\int \mathcal{D}qe^{\frac i\hbar \left(S_0+S_{\hbox{\tiny eff}}\right)}\,,
\label{eq:Gafterintegration}
\ee
\labeld{eq:Gafterintegration}
where the action $S_0$ arises from the original $\cal L$ sans chain terms, specifically
\be
S_0=\int dt\left\{ \frac 12\dot q^2-\left( \omega _0^2
             +\frac12\omega _s^2\right) q^2-\frac \lambda 4q^4\right\} \,.
\label{eq:S0}
\ee
\labeld{eq:S0}
The effective action, $\Seff$, is the result of the integration just described over chain degrees of freedom. It is given by
\be
S_{\hbox{\tiny eff}}
             =\int_0^T\int_0^T\,dt\,ds\,K(|t-s|)q(t)q(s) \,,
\label{eq:Seff}
\ee
\labeld{eq:Seff}
with
\be
K(u)
    =\sum_{n=1,3,\dots }^N\tau_n^2\frac{\cos \Omega _n\left( \frac T2-u\right) }{\Omega _n\sin\left( \frac{\Omega _nT}2\right) }\,.
\label{eq:K}
\ee
\labeld{eq:K}

We shall study the (matrix element of the) propagator in the stationary phase approximation; that is, we focus on the action (including $\Seff$) along the extremal ``classical paths.'' The spectral expansion of the propagator [as a function of \emph{all} variables, i.e., \emph{before} the chain-variable integrations and the division of \Eqref{eq:ExpSeff}] is
\be
G(x_0'',x_1'',\dots,t;x_0',x_1',\dots)=
         \sum_\alpha \Phi_\alpha(x_0'',x_1'',\dots) \exp(-iE_\alpha t) \Phi_\alpha^*(x_0',x_1',\dots) \,,
\label{eq:Genergyexpansion}
\ee
\labeld{eq:Genergyexpansion}
where $\alpha$ is an eigenstate label. This implies that the operations of \Eqref{eq:ExpSeff} (the integrations, \emph{and} the division, which leads to energy-phase corrections) lead to
\be
{\cal G}\left( q'',t;q',0\right) = \exp\left(+i\sum \Omega_n t/2\right)
     \sum_\alpha \phi_\alpha(q'') \exp(-iE_\alpha t) \phi_\alpha^*(q') \,,
\label{eq:G1}
\ee
\labeld{eq:G1}
where
\be
\phi_\alpha(q)\equiv\int \Psi_0(x_1,\dots)^*\Phi_\alpha(q,x_1,\dots) \,dx_1\dots  \,,
\label{eq:phidef}
\ee
\labeld{eq:phidef}
and $\Psi_0$ is the ground state of the chain. Note that $E_\alpha$ in \Eqref{eq:G1} is the total energy of the ring, while $E_C\equiv\sum \Omega_n/2$ is the total ground state energy of the chain. Going to imaginary time, $t=-iT$, \Eqref{eq:G1} becomes
\be
\widetilde{\cal G}\left(q'',T;q',0\right) =
     \sum_\alpha \phi_\alpha(q'') \exp\left[-(E_\alpha-E_{C}) T\right]\phi_\alpha(q')^*
\label{eq:FK}
\ee
\labeld{eq:FK}
This is the starting point for many calculations. The simplest would be to get the ground state energy of the chain by letting $T\to\infty$. For numerical work this turns out to be delicate. What one would do is find a semiclassical expression for the total action in \Eqref{eq:Gafterintegration} and consider its large $T$ limit. Unfortunately $S$ tends to a constant, and the energy emerges from its derivatives. The same occurs for the simple harmonic oscillator. Thus for ${\cal L}_{\hbox{\tiny SHO}}=m\dot x^2/2-m\nu^2x^2/2$, the propagator from $a$ to $b$ in imaginary time $T$ is \bea
\widetilde G_{\hbox{\tiny SHO}}(b,T;a)
   &=&\sqrt{\frac 1{2\pi}\frac{\partial^2S}{\partial a\partial b}}\,
      \exp\left(- S_{\hbox{\tiny classical}}  \right) \nonumber \\
   &=&\sqrt{\frac{m\nu}{2\pi\sinh\nu T}}
    \exp\left(\frac{-m\nu}{2\sinh\nu T}\left[(a^2+b^2)\cosh \nu T -2ab\right] \right)
\label{eq:sho}
\eea
\labeld{eq:sho}
For large $T$ the dominant term in the exponential approaches a constant, so that the energy comes from the prefactor, $\sqrt{\partial^2S/\partial a\partial b}$, which as the explicit form above shows is essentially $1/\sqrt{\sinh\nu T}$, yielding the correct ground state energy, $\nu/2$. This means that one needs extremely accurate values in a second derivative of the action, exponentially small in~$T$. For the action itself this precision was attainable in all cases, and for a single harmonic oscillator the second derivative was also reliable, but for the full ring we did not find that the second derivative had the needed precision.

In any case, evaluating the energy is not our primary goal. We know that the Hamiltonian has eigenstates, but we do not know whether they resemble breathers. We also expect that some states, not necessarily eigenstates, but at the least metastable states, should resemble the classical breather state. This much is guaranteed by semiclassical considerations. Our goal then is to show that the actual eigenstates resemble the breathers.

\begin{figure}
\includegraphics[height=.2\textheight,width=.35\textwidth]{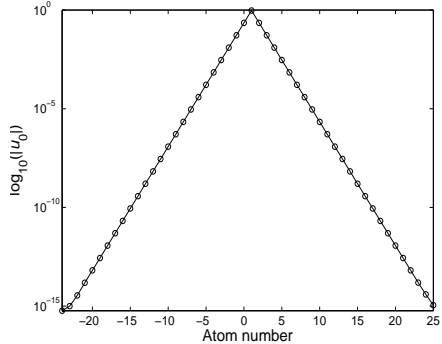}
\caption{The classical local modes are $u_\ell(n)$, with $\ell$ the mode label and $n$ the atom. Shown in the figure is the base-10 logarithm of the absolute value of mode-0, the local mode. Parameters are $\omega_1=2$  and $\omega_0=1$ [cf.\ \Eqref{eq:H01}]. Clearly there is an exponential dropoff of excitation for the atoms.}
\label{fig:localmode}\labeld{fig:localmode}
\end{figure}

To see how to do this, we will consider first the local mode, a quantum state that we know to be localized. Specifically, the classical mode function has a sharp spatial dropoff and the quantum state represents atomic oscillations with that pattern. For reference we show the classical local mode for $\omega_1=2$ and $\omega_0=1$ in Fig.\ \ref{fig:localmode} [a solution of \Eqref{eq:H01}]. (The value of $\omega_s$ does not affect the function.)

Our goal is to ascertain properties of $\phi_\alpha$ and $E_\alpha$ from the semiclassical approximation to $G$. This goal will be attained in Sec.\ \ref{sec:localization}. However, before actually doing this calculation we discuss methodology. Both for the linear and nonlinear cases, implementing the semiclassical approximation presents technical challenges, which we now discuss. We will also indicate the extent to which independent confirmations of our technique are available, taking advantage of known asymptotic properties.

%%%%%%%%%%%%%%%%%%%%%%%%%%%%%%%%%%%%%%%%%%%%%%%%%%%%%
\subsection{Methodology\label{sec:methodology}\labeld{sec:methodology}}
%%%%%%%%%%%%%%%%%%%%%%%%%%%%%%%%%%%%%%%%%%%%%%%%%%%%%%%
The first observation to be made is that the development from \Eqref{eq:lagrangian} to \Eqref{eq:FK} goes through in exactly the same manner, with the same $K$, etc., for the quadratic local mode as for the nonlinear excitation, with the sole exception of the final form of $S_0$, which, for the local mode case, has a term $\omega_1^2q^2/2$ instead of $\lambda q^4/4$. Going over to imaginary time, our problem (both quantum and classical) revolves about the following quantities:
\be 
S_{\hbox{\tiny total}}= \int_0^T dt  \left\{
                   \frac12 \dot q^2 +\frac12 (\omega_s^2 + 2\omega_0^2)q^2 +U(q)\right\}
                   -\int_0^T\int_0^T \,dt\,ds\, \widetilde K(|t-s|)q(t)q(s)   \,,
\label{eq:Scomplex}
\ee 
\labeld{eq:Scomplex}
where $U(q)=\omega_1^2q^2/2$ for the local mode problem. The same form holds for the nonlinear problem as well, but with $U(q)=\lambda q^4/4$. $S_{\hbox{\tiny total}}$ is the total classical action, and in the path integral appears as $\exp(-S_{\hbox{\tiny total}})$ (note the minus sign, a result of the $t\to -it$ transformation). The self-interaction kernel becomes
\be 
\widetilde K(u)=
   \sum_{n=1,3,\dots}^N \tau_n^2
          \frac{\cosh\Omega_n\left(\frac T2-u\right)}
                      {\Omega_n\sinh\left(\frac{\Omega_n T}2\right)}
\label{eq:complexK}
\ee
\labeld{eq:complexK}
The classical equation of motion follows from the usual variational methods and is
\be 
\ddot q-(\omega_s^2+2\omega_0^2)q- \frac{\partial U}{\partial q}       
+2\int_0^T \widetilde K(|t-s|)q(s)=0 \,.
\label{eq:imagtimeclass} 
\ee
\labeld{eq:imagtimeclass} 

For the semiclassical approximation one must solve \Eqref{eq:imagtimeclass}. We used different methods for the linear and nonlinear cases. We discuss them separately.

%%%%%%%%%%%%%%%%%%%%%%%%%%%%%%%%%%%%%%%%%%%%%%%%%%%%%
\subsubsection{Linear, non-local in time propagators\label{sec:linnonlocprop}\labeld{sec:linnonlocprop}}
%%%%%%%%%%%%%%%%%%%%%%%%%%%%%%%%%%%%%%%%%%%%%%%%%%%%%%

For $U(q)=\omega_1^2q^2/2$, \Eqref{eq:imagtimeclass} can be discretized, say with $t_k=k\epsilon$, $k=0,1,\dots,M+1$, $\epsilon=T/M$. $q$ becomes a vector, $q_k=q(t_k)$ ($k=1,\dots,M$) and the entire equation has the form $Bq=q_0$, with $B$ a matrix consisting of three parts: the second derivative operator, $2/\epsilon^2$ on the diagonal, $-1/\epsilon^2$ above and below; a diagonal matrix proportional to $(\omega_s^2+2\omega_0^2)$ and a matrix (which by a slight abuse of notation we call) $K$ whose substantial non-diagonal components reflect the nonlocality in time. From \Eqref{eq:imagtimeclass} it may not be evident how a nonzero right-hand-side ($q_0$) comes into the picture. It arises from the boundary conditions. When discretizing, the first and last rows of the second derivative operator call on $q$-components outside the range $1,\dots,M$. These other components are the boundary values (call them $a=q_\tinybox{initial}$ and $b=q_\tinybox{final}$), so that $q_0(1)=-a/\epsilon^2$ and $q_0(M)=-b/\epsilon^2$ (other components of $q_0$ are zero). This method of dealing with the two-time boundary value problem is a simpler version of what is done in \cite{diffdiff}. It follows that the solution of the nonlocal-in-time equations of motion is $q=B^{-1}q_0$. It is also immediate that for the linear equation the action along this ``classical path'' is given by $S=[q(t)\dot q(t)/2]\big|^T_0$, although one can also evaluate $S$ by explicit integration.

The substantial nonlocality in time means that the matrix $K$ is not sparse, putting a limit on the smallness of $\epsilon$, hence on the accuracy of the action. This difficulty was overcome by the following device. For each such $\epsilon$, call $S(\epsilon)$ the action that results from the discretization and matrix inversion just described. Considered as a function of $\epsilon$, one can write $S(\epsilon)=S(0)+\epsilon S'(0)+\epsilon^2 S''(0)/2 + \dots$, where the prime indicates derivative with respect to $\epsilon$. By evaluating $S(\epsilon)$ for several values of $\epsilon$ one can then extrapolate to zero. For example, the simplest such extrapolation, for two values, $\epsilon_1$ and $\epsilon_2$ (and associated $S_1$ and $S_2$), gives $S(0)=(\epsilon_2S_1-\epsilon_1S_2) /(\epsilon_2-\epsilon_1)$. We typically used three values, chosen as small as possible, but far enough apart to keep denominators from being too small. As we indicate below, this gave excellent results when tested against quantities that could be independently calculated. The following tests were performed:
\begin{itemize}

\item Simple harmonic oscillator. For the energy this requires taking the second derivative with respect to position. This was done by evaluating $G$, hence $S$, at several points and using
\be
\left.\frac{\partial^2 S}{\partial a\partial b}\right|_{a=b}\approx 
\frac{S(a+\delta,a+\delta)+S(a-\delta,a-\delta)-S(a+\delta,a-\delta)-S(a-\delta,a+\delta)}{(2\delta^2)}
\ee
For this formula the extrapolation to $\delta=0$ was also used (as for $\epsilon$ above), which is to say, several values of $\delta$ were used. Furthermore, there was also an extrapolation to $T=\infty$ (by fitting the logarithm of $\partial^2S/\partial a\partial b$ as a function of $1/T$). Results were accurate to better than one part in 10$^4$. 

\item Asymptotics of $\phi_0(q)$ [defined in \Eqref{eq:phidef}]. This can be calculated by non-path integral methods as follows. All eigenstates---both ring and chain---are Gaussians. For the chain the frequencies are $\Omega^{(\tinybox{chain})}_n=\sqrt{\omega_s^2+4\omega_0^2\sin^2(n\pi/2(N+1))}$, $n=1,\dots,N$. For the ring, for the case $\omega_1=0$, they are $\Omega^{(\tinybox{ring})}_n=\sqrt{\omega_s^2+4\omega_0^2\sin^2(n\pi/(N+1))}$, $n=0,\dots,N$. If $\omega_1\neq0$, they are easily evaluated numerically. Aside from normalization, $\phi_0$ is therefore given by [cf.\ \Eqref{eq:phidef}]
\be
\phi_0(q)=\int \prod_{n=1}^N dQ_n^\parentinybox{chain}
           \exp\left(-\textstyle{\sum_k} \Omega^\parentinybox{chain}_k Q_k^{\parentinybox{chain}2}/2\right)
           \exp\left(-\textstyle{\sum_\ell} \Omega^\parentinybox{ring}_\ell Q_\ell^{\parentinybox{ring}2}/2\right)  \,.
\label{eq:phiasgaussian}
\ee
\labeld{eq:phiasgaussian}
Given the transformations from the original variables, $(q, x_1, \dots, x_N)$ to the chain and ring variables (known analytically for the chain, numerically for the ring), it is straightforward to get the coefficient of $q^2$ in $\phi_0$. Let the matrices $U^\parentinybox{ring}$ and $U^\parentinybox{chain}$ be defined by $\Qring{k}= \sum_0^N U^\parentinybox{ring}_{k\ell} x_\ell$ and $\Qchain{k}= \sum_1^N U^\parentinybox{chain}_{k\ell} x_\ell$ (with $\{Q\}$ the appropriate normal coordinates). The $U$'s can be taken to be real. Let 
\bea
u_k    &\equiv& U^\parentinybox{ring}_{k0}\,, \qquad  \Omega_u   \equiv \sum_0^N \Omring{k} u_k^2 \,, \\
g_n    &\equiv& \sum_{k=0}^N \sum_{\ell=1}^N \Omring{k} u_k 
                 U^\parentinybox{ring}_{k\ell}U^\parentinybox{chain}_{\ell n} \,,\\
W_{nm} &\equiv&  \delta_{nm}\Omchain{n} + \sum_{k=0}^N \sum_{\ell=1}^N \sum_{\ell'=1}^N 
                              U^\parentinybox{chain}_{n\ell} U^{\parentinybox{ring}\dagger}_{\ell k}
                              \Qring{k}
                              U^{\parentinybox{ring}}_{k\ell'}U^\parentinybox{chain}_{\ell' m} \,.
\eea
(Note that $U^{\parentinybox{chain}\dagger}=U^\parentinybox{chain}$.) Then the coefficient of $-q^2$ in the exponent of $\phi_0^2$ (call it $\zeta$) is
\be
\zeta=\Omega_u-g^\dagger W^{-1} g \,.
\ee

To evaluate $\zeta$ by the path integral method, the action was evaluated with various endpoints, $a$. Since $-S(a,T;a)$ (the action from $a$ to $a$ in time $T$) appears in the exponent of the propagator, if the semiclassical method and our extrapolations are valid, $S$ should be quadratic in $a$. This was tested by fitting $\sqrt S$ against $a$. We did not do this for a variety of $T$ values since a single large $T$ was sufficient. Large in this case means with respect to the smallest spectral gap, and can be discerned numerically by a stabilizing of the slope in the fit. The desired result is that first, the fit of $\sqrt S$ against $a$ should be a straight line, and second, that the square of the slope of the line should agree with $\zeta$. We will not reproduce the graph of $\sqrt S$ against $a$ since there is little to note beyond its being an excellent fit. The values of slope-squared and $\zeta$ agreed to better than 1\% for $\min(\omega_1,\omega_s)\geq1$ and by no more than 3\% when these quantities became smaller (they impact the spectral gap, which can force $T$ values too large for good accuracy).
\end{itemize}

%%%%%%%%%%%%%%%%%%%%%%%%%%%%%%%%%%%%%%%%%%%%%%%%%%%%%%%%%%%
\subsubsection{Non-linear, non-local in time propagators\label{sec:nonlinnonlocprop}\labeld{sec:nonlinnonlocprop}}
%%%%%%%%%%%%%%%%%%%%%%%%%%%%%%%%%%%%%%%%%%%%%%%%%%%%%%%%

With a nonlinear term in the action, the classical nonlocal two-time boundary value problem cannot be solved by matrix inversion. The equation to be solved is
\be 
\ddot q-(\omega_s^2+2\omega_0^2)q- \lambda q^3
            +2\int_0^T \widetilde K(|t-s|)q(s)=0 \,.
\label{eq:imagtimeclassnonlin} 
\ee
\labeld{eq:imagtimeclassnonlin}
Our method is an extension of Feynman's approach \cite{feynmanpibook} in which the introduction of an auxiliary variable eliminates the nonlocality. Then one can use standard numerical methods for solving the two-time (even-if-nonlinear) boundary value problem. Recalling the definition of $\widetilde K$, \Eqref{eq:complexK}, one can define auxiliary variables $z_n$ by
\be 
z_n(t)\equiv
   \int_0^T \,ds\, \frac{\cosh\Omega_n\left(\frac T2-|t-s|\right)}
                {\Omega_n\sinh\left(\frac{\Omega_n T}2\right)} q(s) \,.
\label{eq:zdef} 
\ee 
\Eqref{eq:imagtimeclassnonlin} becomes 
\be 
\ddot q-(\omega_s^2+2\omega_0^2)q-\lambda q^3 
  +2\sum_n \tau_n^2 z_n=0   \,.
\label{eq:withz} 
\ee
Taking two derivatives of $z_n$ leads to
\be
\frac{d^2z_n}{dt^2} = \Omega_n^2 z_n -2q(t)  \,,\qquad n=1,3,\dots,2\left[\frac{N-1}2\right]+1 \,.
\label{eq:zequation}
\ee
One non-local equation has been replaced by a larger number of local ones. There is but a single pair of boundary conditions: $q(0)=a$, $q(T)=b$. The conditions on $z$ are forced by self-consistency arising from \Eqref{eq:zdef}, which incidentally also imply $z_n(0)=z_n(T)$, for all $n$. It's amusing that with the quartic interaction replaced by a quadratic one, this linear system is equivalent to the set of classical linear equations that apply to the ring \cite{note:check}, with the boundary conditions inherited from the ground state averaging [which is of course where \Eqref{eq:imagtimeclassnonlin} (made linear) comes from in the first place].

For numerical solution of \Eqref{eq:imagtimeclassnonlin} we applied a variation of this method. As a function of its argument, $\widetilde K(u)$, although a sum of many hyperbolic cosines, actually (for large $N$ and $T$) bears a strong resemblance to a single such function. For the parameter range of interest, one can, by judicious choice of $\omega_{\tinybox{eff}}$, bring the difference between $\widetilde K$ and its approximation, $\widetilde K_1(u)\equiv \widetilde K(0)\cosh (\omega_{\tinybox{eff}}u)$, to about $10^{-4}$ (integral of square of difference). With a sum of two hyperbolic cosines (and 3 adjustable parameters) one can do much better, but as we shall see below a single hyperbolic cosine was sufficiently accurate for excellent minimization of the action, i.e., solution of the classical motion.

With a single hyperbolic cosine in the nonlocal portion of the equation, only a single $z$ need be defined [cf.\ the derivation of \Eqref{eq:zequation}] so that our nonlocal equation becomes a pair of second order ODE's, with boundary values for one of them and a self-consistency condition for the boundaries of the other. Using the optimal $\omega_{\tinybox{eff}}$ defined in the last paragraph, the equations are
\bea
 \ddot q &=&  (\omega_s^2+2\omega_0^2)q+\lambda q^3  -2 z   \,,\\
 \ddot z &=& \omeff^2 z -2q(t) \,, \qquad \hbox{with~}
  z(t)\equiv \int_0^T \,ds\, \frac{\cosh\omeff\left(\frac T2-|t-s|\right)}
                {\omeff\sinh\left(\frac{\omeff T}2\right)} q(s)  \,.
\eea
The boundary values of $q(t)$ are given. For $z$ one proceeds iteratively. For given $z(0)$ (which automatically equals its value at $T$), the boundary value problem can solved using {\small MATLAB}'s program ``bvp4c'' \cite{matlabbvp}. Using the associated solution, $q(t)$, one can recompute $z(0)$. In effect one has a function mapping $z(0)$ into a new value, and again, standard numerical search techniques can be use to find a fixed point of this map. That fixed point then provides a solution of the original boundary value problem for~$q$.

In principle with a better approximation for $\widetilde K$, using say two hyperbolic cosines, the parameter space of boundary value mappings becomes 2-dimensional. This, however, was not the way we proceeded.

Since the true solution of the classical nonlinear, nonlocal problem minimizes the action, $S$, it is possible to improve the solution by modifying the $q$ derived from the process given above in such a way as to reduce $S$. The class of functions to add to $q(t)$ for this purpose can be narrowed by the following consideration. The potential that we study is an inverted oscillator, linear or nonlinear. Therefore if the time interval for going between two not-too-small boundary values is large, for most of that time interval the particle will be near zero, with zero velocity: the path is thus an \textit{instanton} and, except near the endpoints, will be exponentially small (in $T$). Profitable, i.e., $S$-reducing, variations of $q$ will thus have the same shape. Our basic variation consists of adding and subtracting hyperbolic cosine functions with varying amplitudes and angular frequencies.

Thus, starting with the approximate-$\widetilde K$ solution to the nonlinear boundary value problem, we allowed modifications of the sort just described. The changes in $S$ that resulted were extremely small. The first such correction was on the order of 10$^{-5}$ of the action, and subsequent reductions were of order 10$^{-15}$. No other functional forms (e.g., multiplying instanton-shaped curves by oscillatory functions) gave any improvement at all. The results presented below all use this method of optimizing~$S$.

\begin{figure}
\includegraphics[height=.4\textheight,width=.7\textwidth]{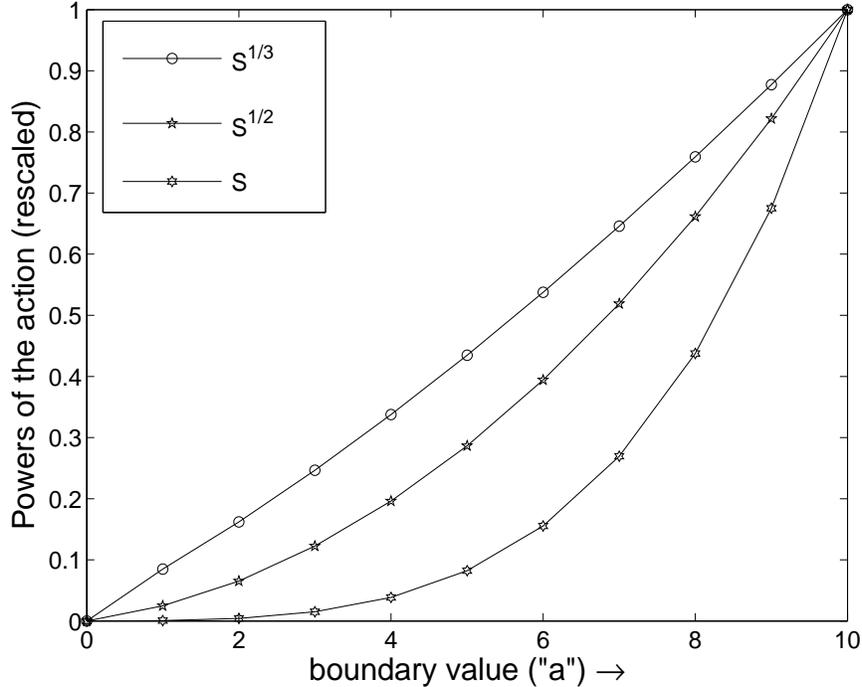}
\caption{Action as a function of boundary value. Since the wave function is essentially $\exp(-S)$, the graph gives the its spatial dropoff. Like the wave function for a single quartic oscillator, the behavior of $\phi_0(q)$ is essentially $\exp(-q^3)$. For this figure, $\lambda=10$ and $T=3$.}
\label{fig:ActionDropoff}\labeld{fig:ActionDropoff}
\end{figure}

It is also possible to perform some of the checks made on the linear problem. In particular the asymptotic form of the wave function for a quartic anharmonic oscillator is $\psi\sim \exp\left(-[\hbox{positive const}\cdot q^3]\right)$. With various $\lambda$ we found the action for large boundary values of $q$. There was a very good fit to the cubic. 
In Fig.\ \ref{fig:ActionDropoff} we plot various powers of the rescaled action, showing that the dropoff for what we have called $\phi_0(q)$ is close to cubic (the exact wave function is also not exactly cubic). 

%%%%%%%%%%%%%%%%%%%%%%%%%%%%%%%%%%%%%%%%%%%%%%%%%%%%%%%%%%
\subsection{Localization\label{sec:localization}\labeld{sec:localization}}
%%%%%%%%%%%%%%%%%%%%%%%%%%%%%%%%%%%%%%%%%%%%%%%%%%%%%%%%%

The local mode represents a localized excitation. Fig.\ \ref{fig:localmode} shows the classical mode oscillation amplitude, hence the shape of the phonon; in particular it indicates an exponential dropoff with distance from the center of the mode. We will use our path integral formalism to show how it too reflects the localization of the ``local mode.'' Then we will use the same technique to establish the localization of the \emph{nonlinear} breather wave function.

Recall the fictitious coupling, $\mu x_0(x_m+x_{N+1-m})$, inserted in the Lagrangian of  \Eqref{eq:lagrangian}. Site-$m$ is far from site-0, where the large oscillations of the breather are taking place. Comparing the action for \emph{small} $\mu$ and the action for \emph{zero} $\mu$ provides a correlation function in the following way. The imaginary-time version of \Eqref{eq:Gafterintegration} is
\be
{\cal G}\left( q^f,T;q^i,0\right) =\int {\cal D}q
              e^{-S_{\tinybox{total}}/\hbar },
\label{eq:Gimaginarytime}
\ee
\labeld{eq:Gimaginarytime}
where there are slight differences from the definitions given in Eqs.\ \parenref{eq:S0} through \parenref{eq:K}. Specifically, the action ($S_{\tinybox{total}}$) is now given by \Eqref{eq:Scomplex}, with $\widetilde K$ as defined there [note that \Eqref{eq:Scomplex} includes both the cases of quadratic and quartic potentials]. We emphasize that the fictitious coupling appears in the definition of $\tau_n$ [\Eqref{eq:tau}], and \emph{only} there. 

Thinking back to the propagator before the integration over chain mode ground states, we consider the derivative $\partial G(a,-iT;a)/\partial \mu|_{\mu=0}$. The importance of this derivative arises from the relation
\be
\langle A\rangle = \frac\partial{\partial \mu}\left[ \int \exp(-S+\mu A)dx\right]\Bigg|_{\mu=0}
                    \Bigg/\int\exp(-S)dx  \,,
\label{eq:correlationdefiniition}
\ee
\labeld{eq:correlationdefiniition}
where $\exp(-S)$ is a weight for averaging. Since we are taking $\partial/\partial \mu$ after integrating, we are getting an average of the $\langle qx_m\rangle$-correlation in the chain mode ground state \cite{note:qxmcorrelation}. 

Note too that because of our use of the semiclassical approximation (which happens to be exact in the linear case), study of $\partial G(a,-iT;a)/\partial \mu|_{\mu=0}$ is essentially the same as study of $\sigma(a,T,g) \equiv \partial S(a,T)/\partial \mu|_{\mu=0}$, where ``$S$'' is the imaginary time action, $a$ is the common initial and final endpoint, and $T$ the imaginary time. The quantity $g$, implicit in  $S(a,T)$, parameterizes the relevant particle-0 enhancement; for the linear local mode it is $\omega_1$ and for the quartic case it is~$\lambda$. 

\begin{figure}
\centerline{
\includegraphics[height=.3\textheight,width=.4\textwidth]{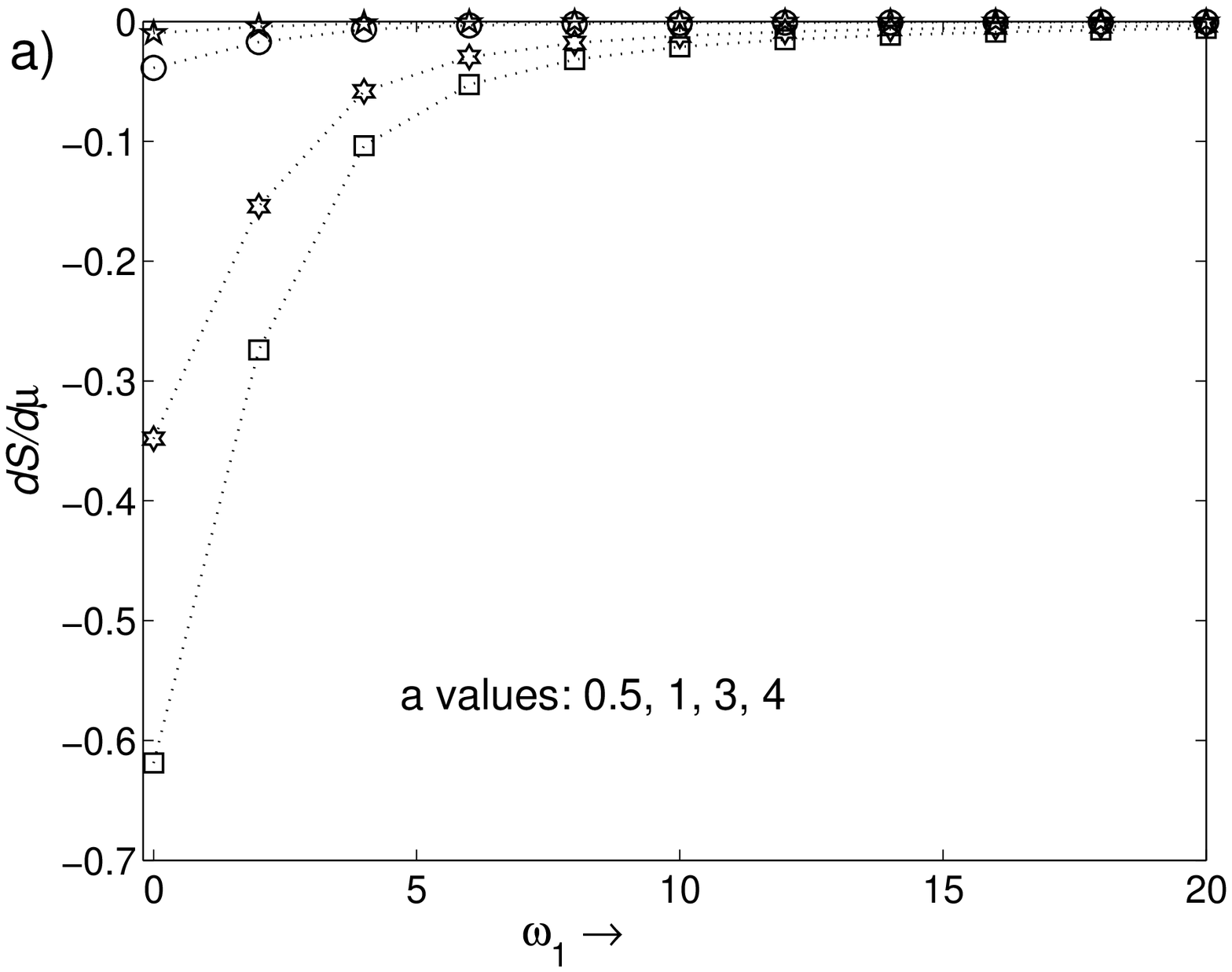}~~
\includegraphics[height=.3\textheight,width=.4\textwidth]{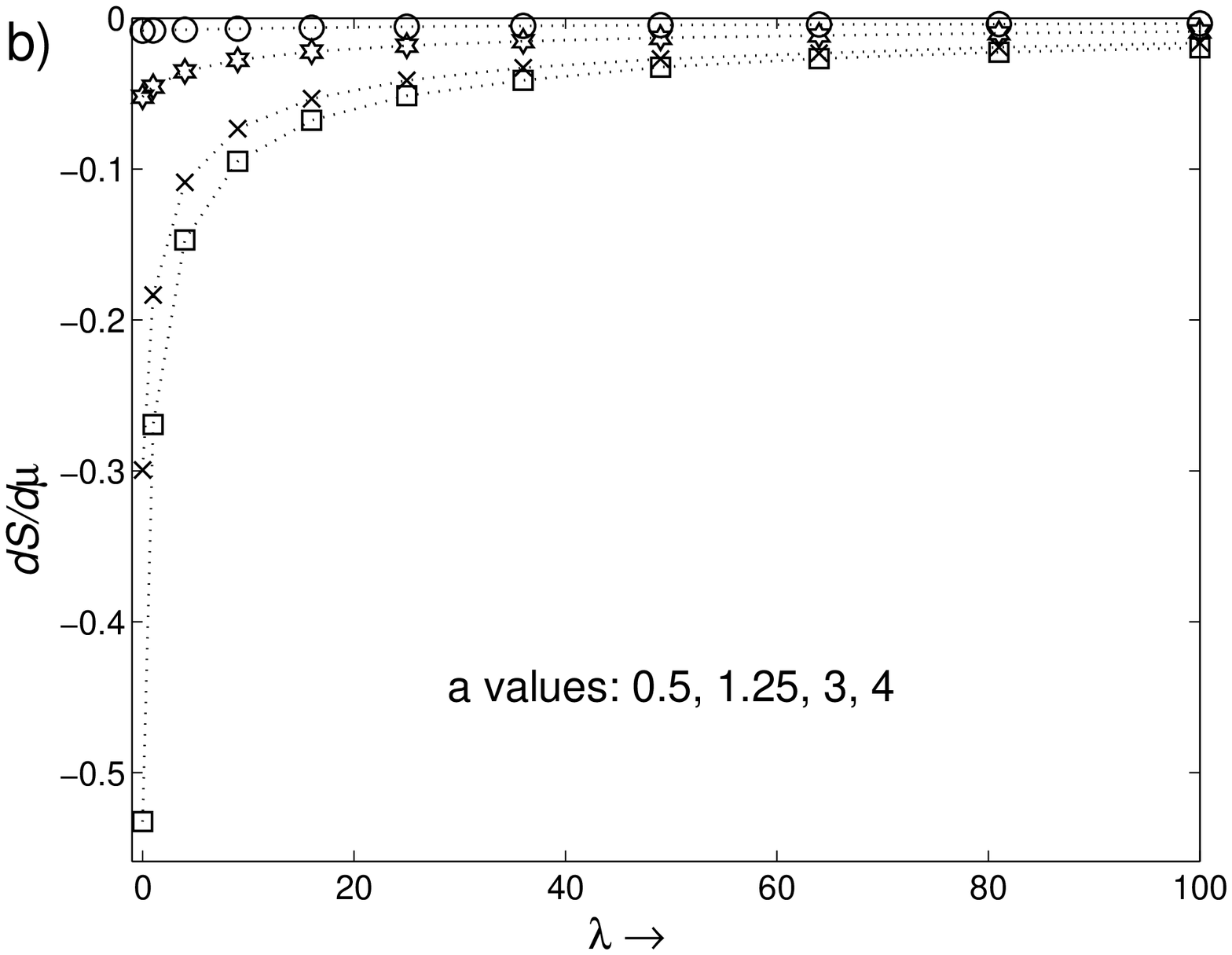}%
}%
\caption{$dS/d\mu$ (essentially a correlation function) has markedly different behavior for large and small endpoint (``$a$'') values for $q$. Both for a linear local mode (as a function of $\omega_1$) and for the breather (function of $\lambda$), a large value of $a$ demands relatively large correlations when neither a local mode nor a breather is present ($\omega_1\sim\lambda\sim0$), but that correlation is wiped out for large parameter ($\lambda$ or $\omega_1$), for which the breather or local mode is effectively decoupled from the rest of the ring.}
\label{fig:vanishcorr}\labeld{fig:vanishcorr}
\end{figure}

Consider first $\sigma(a,T,\omega_1) \equiv \partial S(a,T)/\partial \mu|_{\mu=0}$ for moderate $T$ (such that states other than the ground state survive in the spectral sum for $\cal G$), for small $\omega_1$ and for large $a$. Because $a$ is large, the important terms in the spectral sum will not be those of lowest energy, but those that permit large excursions of $q$ [cf.\ \Eqrefs{eq:G1}{eq:phidef}]. However, since $\omega_1$ is assumed small, there will be no local mode. So pulling $q$ far from its equilibrium position, pulls \emph{all} atoms far from their positions, and the correlation of atom-0 ($q$) and atom-$m$ ($x_m$) should be large. On the other hand, suppose $\omega_1$ to be large (with $a$ still large and $T$ moderate). In that case there is a pronounced local mode and pulling $q$ away from equilibrium has little impact on $x_m$. For the local mode, $q$ can have large excursions while other atoms hardly move. Therefore one expects $\sigma$ to be small. In Fig.\ \ref{fig:vanishcorr}a, the lowest curve shows just this behavior. The boundary value of $q$, $a$, is 4 and for small $\omega_1$ (no local mode) there is a large correlation with the motion of atom-$m$ (0 is at the \emph{top} of the figure). As $\omega_1$ increases, this correlation shrinks.

By contrast, if $a$ is small, the variation of $\omega_1$ has little effect. This can be understood as follows. The requirement on the endpoints of $q$ imposes little demand on any other coordinates whether or not there is a local mode.

\begin{figure}
\centerline{
\includegraphics[height=.3\textheight,width=.4\textwidth]{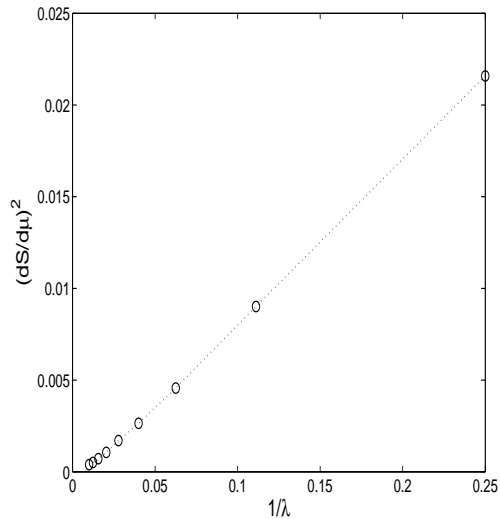}%
}%
\caption{Plot of $(dS/d\mu)^2$ versus $1/\lambda$ for $a=4$. Same data as Fig.\ \ref{fig:vanishcorr}b.}
\label{fig:dSdmuvsoneoverlambda}\labeld{fig:dSdmuvsoneoverlambda}
\end{figure}

We next turn to the nonlinear case. Here we do not have a priori knowledge of the wave function but can use the correlation function as a test of localization. The behavior of the correlation function, as a function of $a$ and $\lambda$ exactly parallels that of the linear local mode. This is shown in Fig.\ \ref{fig:vanishcorr}b. For small $\lambda$ and large $a$ there is no breather and, as for the small $\omega_1$ case, the demand for a large excursion of $q$ forces a large excursion of $x_m$. And now the central observation: for large $\lambda$, forcing $q$ to be large has almost no impact on $x_m$, exactly as for the quadratic local mode, from which we deduce the localization of the breather excitation. We remark that corresponding values of $\omega_1$ and $\lambda$ are related by $\lambda\sim\omega_1^2$. Finally we mention that from Fig.\ \ref{fig:vanishcorr}b alone it is difficult to tell whether $dS/d\mu$ is tending to a constant or to zero. In Fig.\ \ref{fig:dSdmuvsoneoverlambda} is a plot of $(dS/d\mu)^2$ versus $1/\lambda$, in which the extrapolated value is close to zero.

\begin{figure}
\centerline{
\includegraphics[height=.285\textheight,width=.4\textwidth]{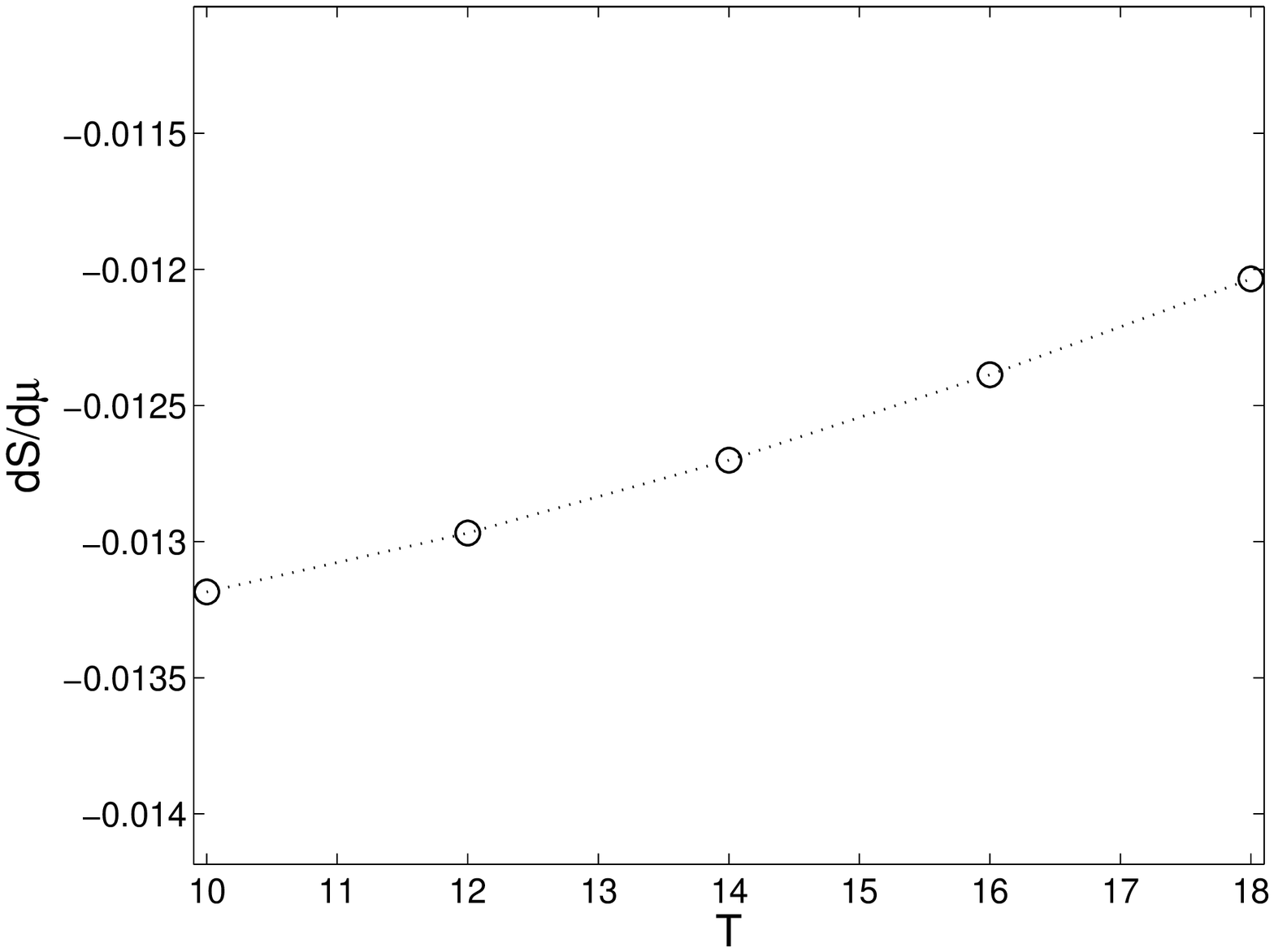}~~
\includegraphics[height=.3\textheight,width=.4\textwidth]{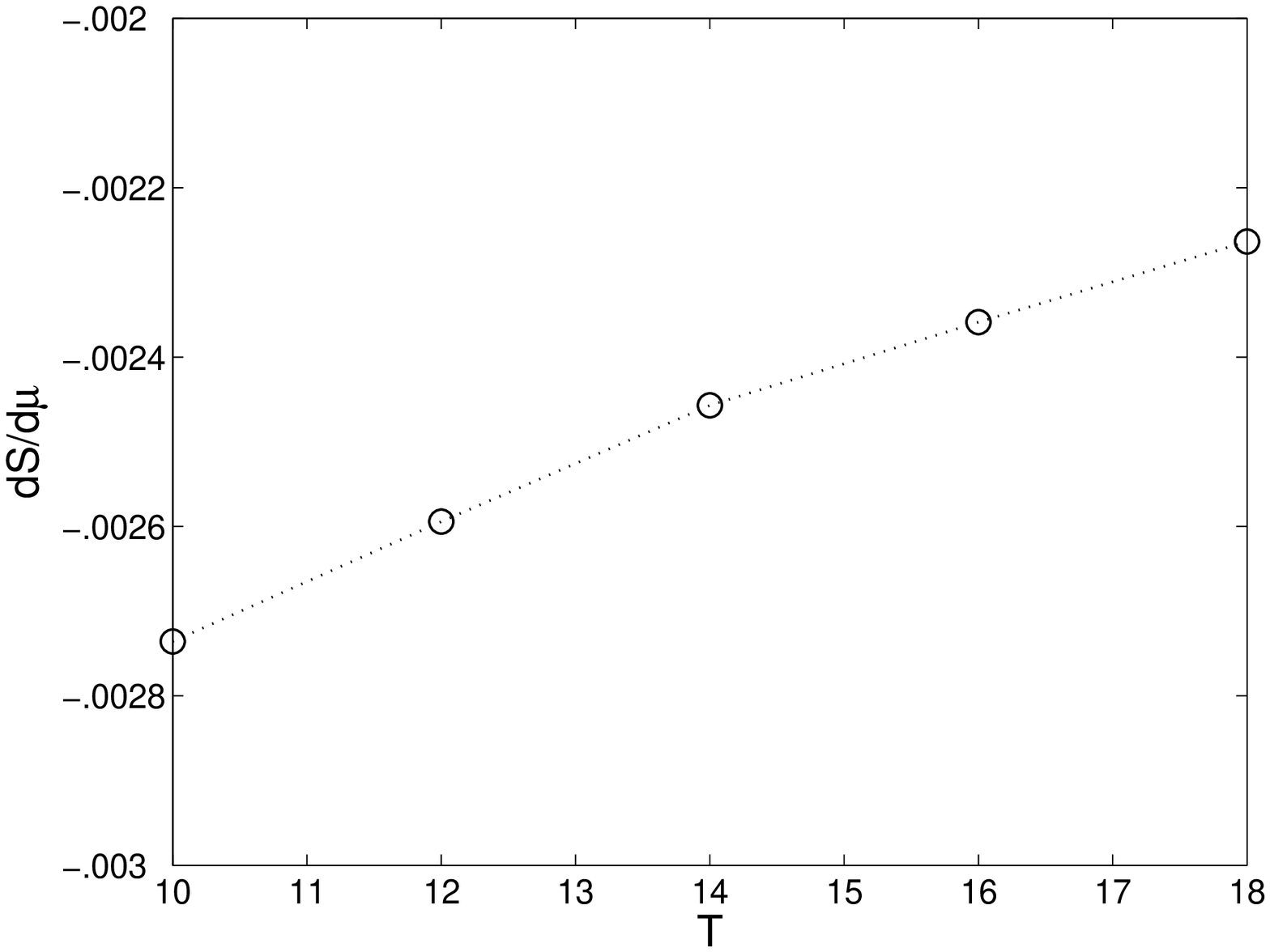}%
}%
\caption{$dS/d\mu$ as a function of time. The figures show a decline with fixed $a$ and coupling parameter ($\omega_1$ or $\lambda$) as $T$ increases. This is due to the increasing weight, with increasing $T$, of the ground state in the propagator. Cf.\ \Eqref{eq:G1}.}
\label{fig:correlationtimedependence}\labeld{fig:fig:correlationtimedependence}
\end{figure}

Another check of correlation function behavior is its time dependence. For a localized stable state and fixed $a$ (the boundary value for $q$) $dS/d\mu$ should decrease with time. This follows from consideration of the propagator given in either \Eqref{eq:Genergyexpansion} or \Eqref{eq:G1}. For large $T$ and $a$ there is a competition between the terms in the expansion, with large $T$ favoring states of lower energy and large $a$ favoring states that drop off most slowly in their spatial coordinate, which in general will be of higher energy. As $T$ increases lower energy states are increasingly important in the mix and the correlation correspondingly reduced (note that for the correlation the overall magnitude of the propagator, and in particular the factor $\exp(-E_0T)$, drops out). In Fig.\ \ref{fig:correlationtimedependence} this decline can be seen for both the quadratic local mode and the quartic breather. On the other hand, if the quantum breather were decaying in time the correlation should \emph{increase}, since the true ground state of the system (if the breather were not stable) would resemble the $\lambda=0$ case, for which the correlation is high.

%%%%%%%%%%%%%%%%%%%%%%%%%%%%%%%%%%%%%%%%%%%%%%%%%%%%%%
\section{Direct diagonalization\label{sec:diagonalization}\labeld{sec:diagonalization}}
%%%%%%%%%%%%%%%%%%%%%%%%%%%%%%%%%%%%%%%%%%%%%%%%%%%%%%

The full Hamiltonian governing our system is given by \Eqref{eq:H1} [or \Eqref{eq:H2}] and it is natural to seek the properties of the system by numerical diagonalization of $H$ in an appropriate basis. Indeed this is the approach of \cite{wang}. The difficulty lies in the fact that a cutoff in phonon number is necessary in order to keep the problem finite. Since the overall Hilbert space is a product of individual phonon Hilbert spaces the total dimension is the product of that of the individual subspaces, so that the size of the matrix representing $H$ is large. Moreover, even when one has the eigenstates of $H$ (which by definition are stable), one still needs to show them to be localized in order to identify them with breathers. (Evidence for localization is also given in \cite{wang}.)

\begin{figure}
\includegraphics[height=.4\textheight,width=.6\textwidth]{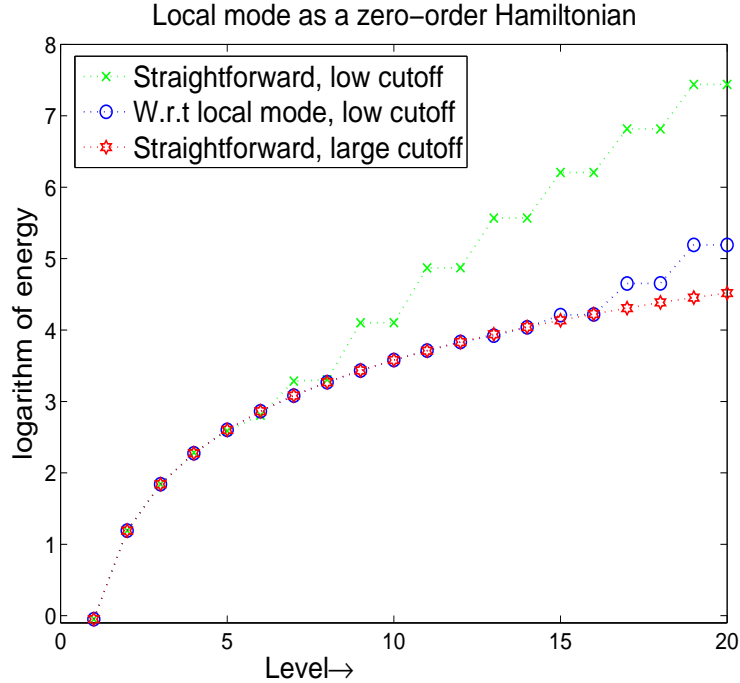}
\caption{Advantage of using the local mode for perturbation theory. Let $H_0=p^2/2+x^2/2$, $V_4=\lambda x^4/4$, $V_2=\omega_1^2x^2/2$, $V_{\tinybox{4 minus 2}}=V_4-V_2$. For this graph we use $\lambda=8$ and $\omega_1=3$. The curve marked ``Straightforward, low cutoff'' results from perturbing $H_0$ by $V_4$, with operators in the number representation cut off at 20. The curve marked ``W.r.t. local\dots'' takes $H_0+V_2$ as the ``zero-order'' Hamiltonian and perturbs it by $V_{\tinybox{4 minus 2}}$ with the same cutoff, 20. This parallels our method for the ring problem. Finally the curve marked ``large cutoff'' results from perturbing $H_0$ by $V_4$ but with a cutoff of 100, yielding reliable results for levels 1 through 20. Note that the graph shows the (natural) logarithm of the energies (with $E_0$ close to 1 for this $\lambda$).}
\label{fig:benefitoflocalmode}\labeld{fig:benefitoflocalmode}
\end{figure}

To reduce the burden of dimensional proliferation we used two devices. First, we introduce a second system for comparison, namely the Hamiltonian \Eqref{eq:H01} [or \Eqref{eq:H02}], representing a local mode. The idea is that if $\omega_1$ (the fictitious self-coupling at site-0) is chosen appropriately the frequency and to some extent the shape of the breather (induced by a quartic with coefficient $\lambda$) can be reasonably approximated. In that way, the perturbation, has minimal effect, and only a small number of non-localized phonon states enter the eigenstate associated with the breather. See Fig.\ \ref{fig:benefitoflocalmode} for an illustration of the efficacy of this method. The second device takes advantage of the reflection symmetry embodied in $P$ [\Eqref{eq:P}]. Since the perturbation ($\lambda x_0^4/4 -\omega_1^2x_0^2/2$) commutes with $P$ we can focus on states in the same symmetry class as the local mode, namely those with $P$-eigenvalue~1. This reduces the number of phonons by almost a factor two, lowering the Hilbert space dimension to a bit more than the square root of what we would otherwise need to consider. Note that using eigenstates of $P$ means that we are not using ``traveling waves'' for the phonons, and indeed one should no longer expect these to be the preferred basis once translational invariance has been dropped.

\begin{figure}
\includegraphics[height=.3\textheight,width=.5\textwidth]{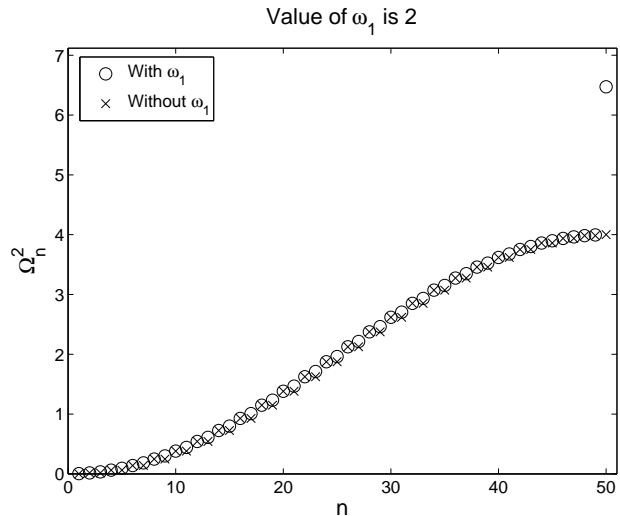}
\caption{Comparing spectra with and without $\omega_1$.}
\label{fig:spectrum}
\end{figure}

Some details of the calculation play a role in interpreting the results and we present them here. The normal modes of our monatomic ring [\Eqref{eq:H01}] satisfy the classical equation of motion,
\be
\ddot x_k +\omega_s^2 x_k +\omega_0^2\left(2x_k-x_{k-1}-x_{k+1}\right)+\delta_{k0}\omega_1^2 x_k  =0 \,,
\qquad k=0,\dots,N+1,
\label{eq:classical}
\ee
\labeld{eq:classical}
with mod-$(N+1)$ addition and $\delta_{jk}$ the Kronecker delta. For $\omega_1=0$, this is trivially solvable, while for non-zero $\omega_1$ there is a local mode. Corresponding equations hold for the nearest-neighbor nonlinear-coupling model [\Eqref{eq:H02}], but since the principles are the same we do not present the equations in detail. With $\omega_0=1$ and $\omega_s$ of order unity, a typical spectrum is close to that of the $\omega_1=0$ case, except for a single mode with frequency high above all the others. An example is shown in Fig.\ \ref{fig:spectrum}.

\begin{figure}
\includegraphics[height=.35\textheight,width=.5\textwidth]{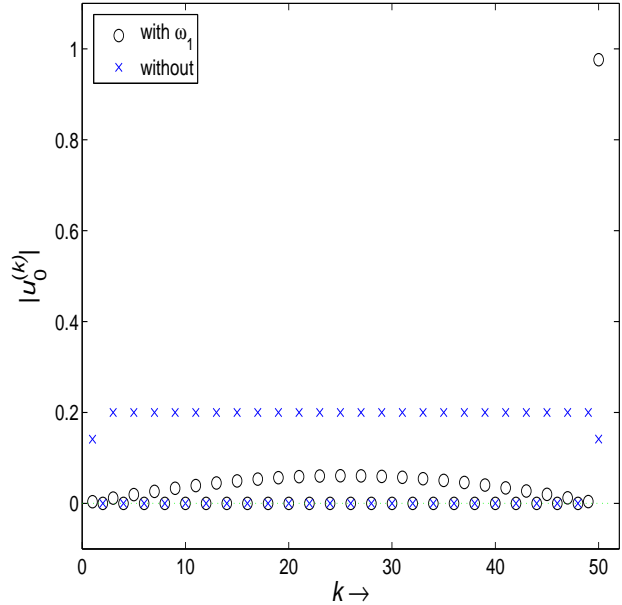}
\caption{Zeroth component of the local mode with and without $\omega_1$.}
\label{fig:firstcomponent}\labeld{fig:firstcomponent}
\end{figure}

At the quantum level the Hamiltonian is written using the classical modes. Call $\Omega_\ell$ the frequency of the $\ell^\tinybox{th}$ mode, $\uupdown{\ell}{}$ the coordinates of the mode [i.e., $\uupdown{\ell}{} \exp(i\Omega_\ell t)$ solves \Eqref{eq:classical}], and let $a_\ell$ be its annihilation operator. The frequencies, $\Omega_\ell$, are not the same as those defined in \Eqref{eq:OmegaChain}; those are for the chain, these for the ring. (When they appear together we distinguish using superscripts [cf.\ \Eqref{eq:phiasgaussian}]). We label the local mode ``0'' (although it is of highest frequency). As usual, $H_0$ becomes
\be
H_0=\sum \Omega_\ell a_\ell^\dagger a_\ell \,,
\label{eq:H0phonon}
\ee
\labeld{eq:H0phonon}
where we now restrict ourselves to modes symmetric under $P$, and call $N_s$ the number of such modes. The perturbation, given earlier, is
\be
V=\quartertext\lambda x_0^4-\htx\omone^2 x_0^2 \,,
\label{eq:V}
\ee
\labeld{eq:V}
so that all we need in order to proceed is the fact that
\be
x_0=\sum\frac1{\sqrt{2\Omega_\ell}}\left(a_\ell+a_\ell^\dagger\right)\uupdown{\ell}{0}
\,.
\label{eq:x1}
\ee
\labeld{eq:x1}
Note that it is the \textit{zeroth} component of each mode function, $\uupdown{\ell}{}$, that is important. (Had the asymmetric states been included they would now drop out since their zeroth components vanish.) In our general discussion we indicated that choosing to perturb around the local mode lessens the impact of the quartic term. In \Eqref{eq:x1} the specific mechanism of that effect can be seen: it is built into the function $\uupdown{\ell}{0}$, which is plotted in Fig.\ \ref{fig:firstcomponent}. Note that without $\omega_1$ (i.e., no local mode) the values of $\uupdown{\ell}{0}$ fluctuate around 0.2 ($u$ is square normalized), while with a moderate $\omega_1$ all but $\uupdown{0}{0}$ drop to much smaller values. 

With these definitions and observations the operator $H$ can be generated. It is a sparse matrix and selected eigenvalues and eigenvectors can be obtained for quite large dimension. 

Our objective is to find the eigenstates corresponding to breathers. The most convenient for our purposes is the first excited state, since it presumably does not have enough excitation for the cutoff to be sensed (cf.\ Fig.\ \ref{fig:benefitoflocalmode}). As a state around which to perform the perturbation we use $|0,\dots,0,1\rangle$, which is the eigenstate of the local mode Hamiltonian, $H_0$ of \Eqref{eq:H01}, having a single excitation of the local mode, with all other phonons in their ground state. The values of the parameters that we use are $\lambda=8$, $\omega_0=1$, and $\oms=1$, which after adjusting for differences of convention, are the values used in \cite{wang}. For $\omone$ we used 2.5, although the structure of the eigenfunction was not sensitive to this. In Table \ref{tab:state} we show the $H_0$-eigenstates with the largest components in the true eigenstate. Clearly the local mode dominates. The next largest component is in fact the thrice excited local mode, which does not represent a spreading of amplitude, but rather a readjustment of the shape of the excitation. Other modes barely make to the $10^{-3}$ level. Of particular interest is the observation that the highest excitation level for other phonons is 3, indicating that a cutoff of 6 is safe. In fact, even to probability levels of $10^{-8}$ there is no excitation higher than 3 for anything but the local mode.

\begin{table}
\caption{Principal components of the breather state. The first 4 columns refer to the 4 symmetric phonons in a 6-atom ring. First row: frequencies. Subsequent rows: number operator values. (The highest frequency is the local mode.) Fifth column: norm squared of the mode (power of 10 multiplier in parentheses). The local mode cutoff was 13, for the others 6. For this state first order perturbation theory is good to 0.3\%. The last column reports the same calculation with cutoff~8.\\ }
\label{tab:state}
\begin{tabular}{|c|c|c|c||c|c|}
\colrule
0.749  &   0.987  &    1.19 &     2.04 & Prob.\ (6) & Prob.\ (8) \\ \colrule
\colrule
0 & 0 & 0 & 1 & 0.9947    &  0.9952    \\   \colrule
0 & 0 & 0 & 3 & 3.315 (-3)&~3.321 (-3) \\   \colrule
0 & 1 & 0 & 0 & 1.122 (-3)& 0.848 (-3) \\   \colrule
0 & 0 & 1 & 0 & 5.748 (-4)& 4.345  (-4) \\   \colrule
1 & 0 & 0 & 0 & 1.545 (-4)& 1.167 (-4) \\   \colrule
0 & 1 & 0 & 2 & 1.004 (-4)& 0.754 (-4) \\   \colrule
1 & 0 & 0 & 2 & 2.866 (-5)& 2.152 (-5) \\   \colrule
0 & 0 & 1 & 2 & 2.303 (-5)& 1.728 (-5) \\   \colrule
0 & 0 & 0 & 5 & 3.670 (-6)& 3.627 (-6) \\   \colrule
\colrule
\end{tabular}
\end{table}

As a check of cutoff sensitivity we repeated this calculation with a cutoff of 8 (but with the local mode still at 13). The results are in the last column of Table \ref{tab:state}. There is little sensitivity to the change---not only in the probabilities but in the composition of the state. 

We also studied larger rings, but with smaller cutoffs. For eight atoms there were 5 phonon modes, with a cutoff of 7 for the local-mode phonon and 5 for the others. Results with $\omega_1=2.3$ are shown in Table \ref{tab:state2}. Once again the principal change from the unperturbed state arises from the excited breather state. And again, down to $10^{-6}$ probability no non-local-mode phonon has excitation greater than one, an indication that the cutoff is again not felt. It is interesting that in other runs with different values of $\omega_1$ most of the change in the state was in the local-mode-phonon's contribution, which makes sense since it may be a better or worse approximation to the true nonlinear state, depending on~$\omega_1$.

\begin{table}
\caption{Principal components of the breather state. The first 5 columns refer to the 5 symmetric phonons in a 8-atom ring. First row: frequencies. Subsequent rows: number operator values. (The highest frequency is the local mode.) The sixth column is the norm squared of the mode. The local mode cutoff was 7, for the others 5. For this state first order perturbation theory is good to 0.3\%.\\ }
\label{tab:state2}
\begin{tabular}{|c|c|c|c|c||c|}
\colrule
0.73131 & 0.8884 & 1.082 & 1.208 & 1.921 & Prob.\ \\ \colrule
\colrule
0 & 0 & 0 & 0 & 1 & 0.9927    \\   \colrule
0 & 0 & 0 & 0 & 3 & 3.20 (-3) \\   \colrule
0 & 0 & 1 & 0 & 0 & 2.05 (-3) \\   \colrule
0 & 1 & 0 & 0 & 0 & 1.05 (-3) \\   \colrule
0 & 0 & 0 & 1 & 0 & 6.32 (-4) \\   \colrule
0 & 0 & 0 & 0 & 5 & 2.010 (-4) \\   \colrule
1 & 0 & 0 & 0 & 0 & 1.20 (-4) \\   \colrule
0 & 1 & 0 & 0 & 2 & 2.39 (-5) \\   \colrule
0 & 0 & 1 & 0 & 2 & 1.71 (-5) \\   \colrule
0 & 0 & 1 & 0 & 4 & 6.88 (-6) \\   \colrule
0 & 1 & 0 & 0 & 4 & 6.85 (-6) \\   \colrule
1 & 0 & 0 & 0 & 2 & 5.07 (-6) \\   \colrule
0 & 0 & 0 & 1 & 2 & 2.01 (-6) \\   \colrule
0 & 0 & 0 & 1 & 4 & 1.21 (-6) \\   \colrule
1 & 0 & 0 & 0 & 4 & 1.20 (-6) \\   \colrule
\colrule
\end{tabular}
\end{table}

Moving on to 12 atoms, the number of independent phonons is 7 and a cutoff of 3 was imposed on all but the local mode, for which the cutoff was 5. The results, in Table \ref{tab:state3}, continue to be consistent with our previous conclusions.

\begin{table}
\caption{Principal components of the breather state. The first 7 columns refer to the 7 symmetric phonons in a 12-atom ring. First row: frequencies. Subsequent rows: number operator values. (The highest frequency is the local mode.) The eighth column is the norm squared of the mode. The local mode cutoff was 5, for the others 3. For this state first order perturbation theory is good to 0.2\%.\\ }
\label{tab:state3}
\begin{tabular}{|c|c|c|c|c|c|c||c|}
\colrule
0.718 & 0.800 & 0.926& 1.06 & 1.16 & 1.22 & 1.98 & Prob.\ \\ \colrule
\colrule
0 & 0 & 0 & 0 & 0 & 0 & 1 & 0.98995    \\   \colrule
0 & 0 & 0 & 0 & 0 & 0 & 3 & 3.49 (-3) \\   \colrule
0 & 0 & 0 & 1 & 0 & 0 & 0 & 2.25 (-3) \\   \colrule
0 & 0 & 0 & 0 & 1 & 0 & 0 & 1.85 (-3) \\   \colrule
0 & 0 & 1 & 0 & 0 & 0 & 0 & 1.46 (-3) \\   \colrule
0 & 1 & 0 & 0 & 0 & 0 & 0 & 5.55 (-4) \\   \colrule
0 & 0 & 0 & 0 & 0 & 1 & 0 & 3.27 (-4) \\   \colrule
1 & 0 & 0 & 0 & 0 & 0 & 0 & 6.16 (-5) \\   \colrule
0 & 0 & 1 & 0 & 0 & 0 & 4 & 1.76 (-5) \\   \colrule
0 & 0 & 0 & 1 & 0 & 0 & 4 & 1.71 (-5) \\   \colrule
\colrule
\end{tabular}
\end{table}

Of particular interest for our own application is the case of nonlinearity in the nearest neighbor interaction. The results of this calculation support what we have already seen: the breather is very well approximated by phonons of the (quadratic) local mode. In Tables \ref{tab:nbrstate1} and \ref{tab:nbrstate2} we show a small variation on the material displayed for the self interaction. Both the singly excited breather (more precisely, the true eigenstate closest to the singly excited local mode) and the ground state are shown. A variety of $\lambda$ and $\omega_1$ values are also used, to give a richer idea of variation of the state with changes in parameters.

\begin{table}
\caption{Principal components of the breather state with nearest neighbor nonlinearity. Run parameters are followed by a description of the state according to the scheme of the previous tables. For each parameter set \emph{two} states are described, the ground state and the state of a single breather excitation.\\}
\label{tab:nbrstate1}
\begin{tabular}{|c|c|c|c||l|l|}
\hline
\multicolumn{6}{|c|}{$\bf{\omega_{1}}$\bf{=1}  $\;$   $\bf{\omega_{s}}$\bf{=1} $\;$  $\bf{\omega_{0}}$\bf{=1} $\;$ $\bf{\lambda}$\bf{=1}} \\ \hline
\protect\bf { 0.7071}  &   \protect\bf {0.8797}  &  \protect\bf {  1.172} & \protect\bf {1.689} & \protect\bf {Prob.\ (6)} & \protect\bf {Prob.\ (8)}   \\   \hline\hline 
0 & 0 & 0 & 0 & 0.99874   &  0.99875    \\  
0 & 0 & 0 & 2 & 1.184 (-3)&  1.188 (-3) \\   
0 & 0 & 0 & 4 & 3.017 (-5)&  3.012 (-5) \\   
\hline
0 & 0 & 0 & 1 & 0.99711   & 0.9975     \\   
0 & 0 & 0 & 3 & 1.352 (-3)& 1.143 (-3) \\   
0 & 0 & 1 & 0 & 1.137 (-3)& 1.023 (-3) \\   
0 & 1 & 0 & 0 & 2.521 (-4)& 1.905 (-4) \\   
0 & 0 & 0 & 5 & 1.364 (-4)& 1.363 (-4) \\   
\hline\hline
\multicolumn{6}{|c|}{$\bf{\omega_{1}}$\bf{=2}  $\;$   $\bf{\omega_{s}}$\bf{=1} $\;$  $\bf{\omega_{0}}$\bf{=1} $\;$ $\bf{\lambda}$\bf{=1}} \\ \hline
\protect\bf {0.7071} & \protect\bf {0.8832} & \protect\bf {1.175} & \protect\bf {2.709} & \protect\bf {Prob.\ (6)} & \protect\bf {Prob.\ (8)} \\  \hline\hline
0 & 0 & 0 & 0 & 0.96871   & 0.96875    \\   
0 & 0 & 0 & 2 & 3.047 (-2)& 3.047 (-2) \\   
0 & 0 & 0 & 4 & 6.724 (-4)& 6.722 (-4) \\   
\hline
0 & 0 & 0 & 1 & 0.92976   & 0.93058    \\   
0 & 0 & 0 & 3 & 6.538 (-2)& 6.543 (-2) \\   
0 & 0 & 1 & 0 & 2.525 (-3)& 1.901 (-3) \\   
0 & 0 & 0 & 5 & 1.364 (-3)& 1.365 (-3) \\   
0 & 1 & 0 & 0 & 9,504 (-4)& 7.149 (-4) \\   
\hline\hline
\multicolumn{6}{|c|}{$\bf{\omega_{1}}$\bf{=2}  $\;$   $\bf{\omega_{s}}$\bf{=1} $\;$  $\bf{\omega_{0}}$\bf{=1} $\;$ $\bf{\lambda}$\bf{=4}} \\ \hline
\protect\bf { 0.7071}  &   \protect\bf {0.8832}  &  \protect\bf {1.175} & \protect\bf {2.709} & \protect\bf {Prob.\ (6)} & \protect\bf {Prob.\ (8)} \\  
\hline\hline
0 & 0 & 0 & 0 & 0.99072   & 0.99073   \\   
0 & 0 & 0 & 2 & 9.057 (-3)& 9.060 (-3)\\   
0 & 0 & 0 & 4 & 1.842 (-4)& 1.841 (-4) \\   
\hline
0 & 0 & 0 & 1 & 0.99204   & 0.99213 \\   
0 & 0 & 0 & 3 & 6.863 (-3)& 6.865 (-3) \\   
0 & 0 & 0 & 5 & 7.384 (-4)& 7.383 (-4)\\   
0 & 0 & 1 & 0 & 2.295 (-4)& 1.723 (-4) \\   
\hline
\end{tabular}
\end{table}
\begin{table}
\caption{Continuation of Table \ref{tab:nbrstate1}.\\}
\label{tab:nbrstate2}
\begin{tabular}{|c|c|c|c||l|l|}  
\hline
\multicolumn{6}{|c|}{$\bf{\omega_{1}}$\bf{=3}  $\;$   $\bf{\omega_{s}}$\bf{=1} $\;$  $\bf{\omega_{0}}$\bf{=1} $\;$ $\bf{\lambda}$\bf{=3}} \\ \hline
\protect\bf { 0.7071} & \protect\bf {0.8841} & \protect\bf {1.176} & \protect\bf {3.852} & \protect\bf {Prob.\ (6)} & \protect\bf {Prob.\ (8)} \\ \hline\hline 
0 & 0 & 0 & 0 & 0.87366   & 0.87373   \\ 
0 & 0 & 0 & 2 & 0.11745   & 0.11744   \\   
0 & 0 & 0 & 4 & 8.653 (-3)& 8.652 (-3) \\   
0 & 1 & 0 & 1 & 1.116 (-4)& 8.368 (-5) \\   
\hline
0 & 0 & 0 & 1 & 0.80851   & 0.81131 \\  
0 & 0 & 0 & 3 & 0.16949   & 0.17004 \\ 
0 & 0 & 1 & 0 & 1.055 (-2)& 8.022 (-3)\\  
0 & 0 & 0 & 5 & 7.953 (-3)& 7.978 (-3) \\   
0 & 1 & 0 & 0 & 2.336 (-3)& 1.767 (-3) \\   
0 & 0 & 1 & 2 & 9.796 (-4)& 7.463 (-4) \\   
\hline\hline
\multicolumn{6}{|c|}{$\bf{\omega_{1}}$\bf{=3}  $\;$   $\bf{\omega_{s}}$\bf{=1} $\;$  $\bf{\omega_{0}}$\bf{=1} $\;$ $\bf{\lambda}$\bf{=4}} \\ \hline
\protect\bf { 0.7071} & \protect\bf {0.8841} & \protect\bf {1.176} & \protect\bf {3.852} & \protect\bf {Prob.\ (6)} & \protect\bf {Prob.\ (8)} \\ 
\hline\hline
0 & 0 & 0 & 0 & 0.90863   & 0.90867   \\   
0 & 0 & 0 & 2 & 8.784 (-2) & 8.784 (-2)\\   
0 & 0 & 0 & 4 & 3.385 (-3)& 3.385 (-3) \\   
0 & 1 & 0 & 1 & 7.378 (-5)& 5.533 (-5) \\   
\hline
0 & 0 & 0 & 1 & 0.86889   & 0.86957 \\   
0 & 0 & 0 & 3 & 0.12459   & 0.12467 \\   
0 & 0 & 0 & 5 & 3.403 (-3)& 3.405 (-3) \\   
0 & 0 & 1 & 0 & 2.129 (-3)& 1.603 (-3) \\   
0 & 1 & 0 & 0 & 8.827 (-4)& 6.638 (-4) \\   
\hline
\end{tabular}
\end{table}

We summarize: in all cases the (quadratic) local mode dominates the true eigenstate of the Hamiltonian, indicating, because that (quadratic) local mode is itself localized, that the stable eigenstates, or at least those generated by perturbation around the localized modes, are localized. Matching all these properties, including frequencies, we conclude that the quantized breather is stable.

%%%%%%%%%%%%%%%%%%%%%%%%%%%%%%%%%%%%%%%%%%%%%%%%%%%%%%%%%%%%
\section{Time evolution\label{sec:timeevolution}}
%%%%%%%%%%%%%%%%%%%%%%%%%%%%%%%%%%%%%%%%%%%%%%%%%%%%%%%

%%%%%%%%%%%%%%%%%%%%%%%%%%%%%%%%%%%%%%%%%%%%%%%%%%%%%%%%
\subsection{General discussion of decay\label{sec:generaldecay}}
%%%%%%%%%%%%%%%%%%%%%%%%%%%%%%%%%%%%%%%%%%%%%%%%%%

Although we have shown the breather eigenfunction (of the full Hamiltonian) to be dominated by the local modes, nevertheless there were small---very small---contributions from ordinary phonon states (in the local mode basis). Does this imply that a system initially in a state close to the eigenfunction (for example, in a local mode phonon eigenstate), must ultimately decay? An argument in favor of this perspective would be that so long as the initial state has matrix elements that connect it to a continuum of global phonons, it must have a decay rate, simply by Fermi's Golden Rule. And a decay rate implies exponential decay, even if the multiplier of time in that exponent is small.

As a general guide the Golden Rule has wide applicability, but it should not be made into a false idol (a Golden Calf). Several years ago, in response to anomalies in computer decay studies, one of the present authors in collaboration with B. Gaveau \cite{limited} studied one kind of breakdown. The context is a state that has energy $h$ and is coupled to a continuum with energies in the interval $[E_1,E_2]$, with $E_1<h<E_2\leq\infty$. The level with energy $h$ is within the interval and has nonzero coupling under the full Hamiltonian to the other levels. It was found that depending on the threshold behavior of the coupling, its strength, and the proximity of $h$ to an edge, the system need \textit{not} decay exponentially. What can happen is that when the interaction is turned on the initial state loses a bit of amplitude to other modes, but it eventually settles into an asymptotic state of norm well away from zero. To be precise, let the initial state (of energy $h$) be $|0\rangle$ and the full Hamiltonian be $H$. The survival probability is then $p(t)\equiv|\langle0|\exp(-iHt/\hbar)|0\rangle|^2$. Asymptotically, $p(\infty)$ (or a smoothed average) could be anywhere in $(0,1)$.

In \cite{limited} the emphasis was on the mathematical mechanism, namely the formation of what could be called a \textit{plasmon} mode \cite{fano}. Suppose a single mode is coupled to $N$ levels, so that neither that mode nor the $N$ levels are eigenfunctions of the full Hamiltonian. In the usual decay scenario, when that single mode is expanded in eigenstates of the full Hamiltonian its amplitude in each is of order $1/\sqrt N$. Evolving in time, if you wait long enough there is ``nothing'' left in the original state; in the limit $N\to\infty$ the decay is complete. But it can also happen, as it does for plasmons and for the systems studied in \cite{limited}, that the original state has significant [\textit{not} O($1/\sqrt N$)] overlap with a true eigenstate, and as a result only decays to a value dependent on the value of that overlap.

We illustrate both the gradual and catastrophic failure of the Golden Rule for the $(2N+2)\times(2N+2)$ Hamiltonian 
\be
H=\pmatrix{h           &   C^\dagger \cr
           C   &   \Omega \cr}
\label{eq:decaymodel}
\ee
\labeld{eq:decaymodel}
where $C$ is a column vector of coupling coefficients, $h\in \mathbb R$, and $\Omega= \hbox{diag}\, (\omega_{_{-N}}, \dots\omega_{_N})$, with $\omega_n\in[E_1,E_2]$ ($\hbar=1$). (Every decay system can be brought to this form.) Since threshold behavior will prove to be our main preoccupation, behavior at the other end of the energy range is irrelevant. For convenience we take the energy range to be $[-E/2,E/2]$ and allow $h$ to approach $E/2$ from below. The $\omega_n$ are taken to be uniformly spaced---what matters for decay (also when going beyond the Golden Rule) is the product $\rho|\gamma|^2$ with $\rho$ the density of states and $\gamma=C/\sqrt{N}$, which is the appropriate scaling with $N$. Therefore we can make $\rho$ a constant [$(2N+1)/E$], and build dimensional or other density of states features into~$C$. We shall take $C$ of the form $C\equiv(c/\sqrt N)\phi$, with $c$ a constant and $\phi$ of the form $\phi_n=[1-(n/(N+1))^2]^\delta$, with $n=-N,\dots,N$, and $\delta$ another constant. If $h$ is placed near level-$n$ (with its associated $\omega=\omega_n$), the Golden Rule prediction for the decay rate is $\GamGR=2\pi\rho(\omega)|\gamma(\omega)|^2$, which in this case becomes $\GamGR=2\pi(2/E)|c|^2 |\phi_n|^2$ \cite{note:strictly}.

\begin{figure}
\includegraphics[height=.25\textheight,width=.35\textwidth]{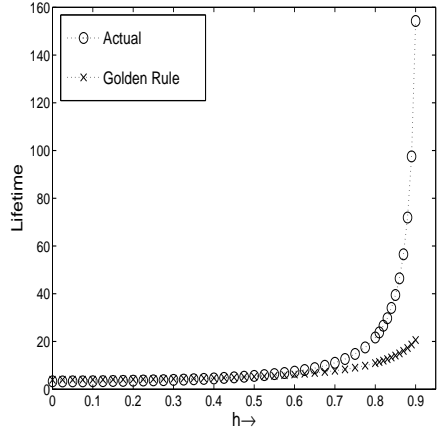}
\caption{``Measured'' lifetime (from computer runs) compared to the Golden Rule prediction as the energy parameter $h$ approaches the band edge (which is $h=1$ here). At $h=0.92$ a plasmon forms and ``lifetime'' loses meaning. Parameter values were $E=2$, $c=0.2$, $\delta=1/2$, $N=200$ for $h\ltsim 0.85$, then increasing values until $N=500$ (matrix size is $2N+2$). Each run was long enough for $p$ to reach $\sim10^{-4}$ (or smaller) with the $\log p$-vs.-$t$ plot remaining a straight line.}
\label{fig:lifetimevsh}\labeld{fig:lifetimevsh}
\end{figure}

\begin{figure}\centerline{
\includegraphics[height=.25\textheight,width=.35\textwidth]{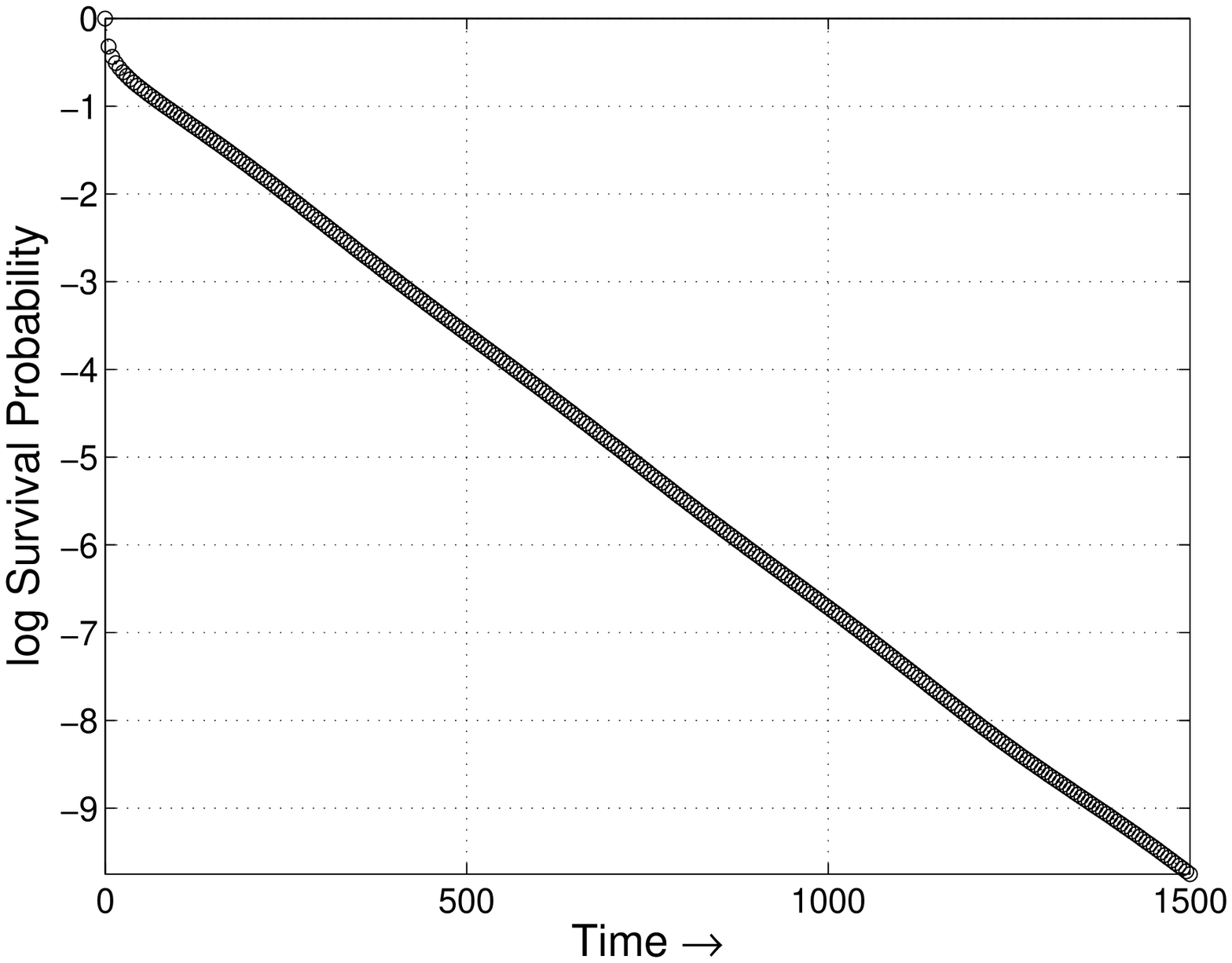}
\includegraphics[height=.25\textheight,width=.35\textwidth]{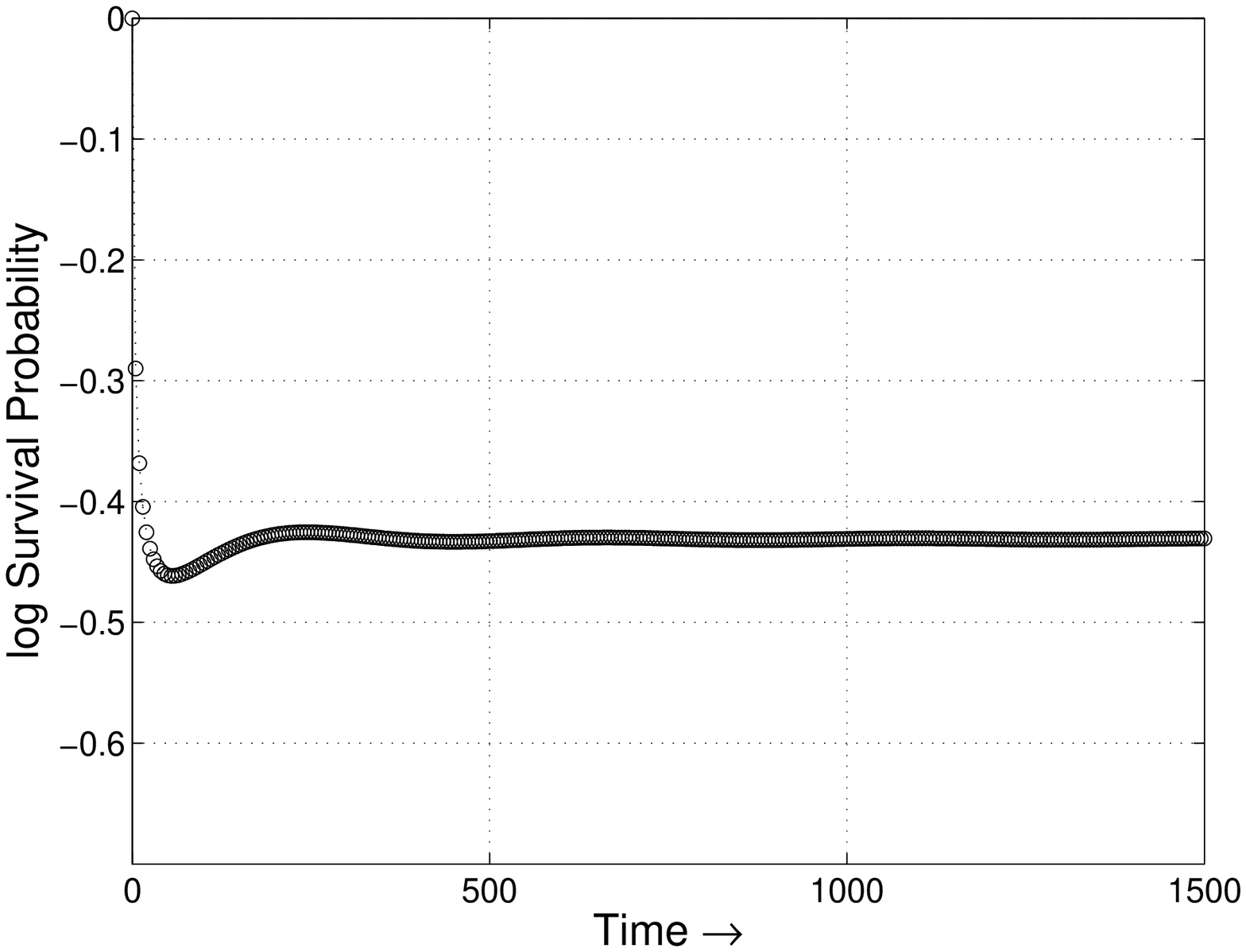}}
\caption{Comparing decay without and with a plasmon ($h=0.90$ and $0.94$, respectively). Note the different $y$-axes ranges. The plasmon energy is only slightly outside the band: the band edge is at 1 and the plasmon energy is 1.016.}
\label{fig:plasmoncomparison}\labeld{fig:plasmoncomparison}
\end{figure}

In Fig.\ \ref{fig:lifetimevsh} we show how the Golden rule prediction gradually declines in accuracy as $h\to E/2$. For all points plotted, there is exponential decay to great accuracy and to values of $p$ that reach 10$^{-4}$ or less. (As lifetimes increase, $N$ is also increased to allow the calculation to go to long times and avoid a Poincar\'e recurrence. For the same $c$, $\delta$ and $h$, changing $N$ had little effect. More on this later.) Finally, for the parameter values stated in the figure, at about $h=0.92E/2$ a \textit{plasmon} mode develops. At this point exponential decay is completely lost. In Fig.\ \ref{fig:plasmoncomparison} we show decay behavior for $h=0.9E/2$, side-by-side with the corresponding graph for $h=0.94$, when a plasmon exists. With the formation of this excitation, a significant fraction of the initial amplitude \textit{never} decays. This has nothing to do with Poincar\'e recurrence. The same asymptotic probability as well as overall pattern of the curve obtains whether we did the calculation with dimension 102 or 1002  ($N=50$ or 500, and several values in between). This is the continuum behavior calculated in \cite{limited}.

\begin{figure}\centerline{
\includegraphics[height=.25\textheight,width=.35\textwidth]{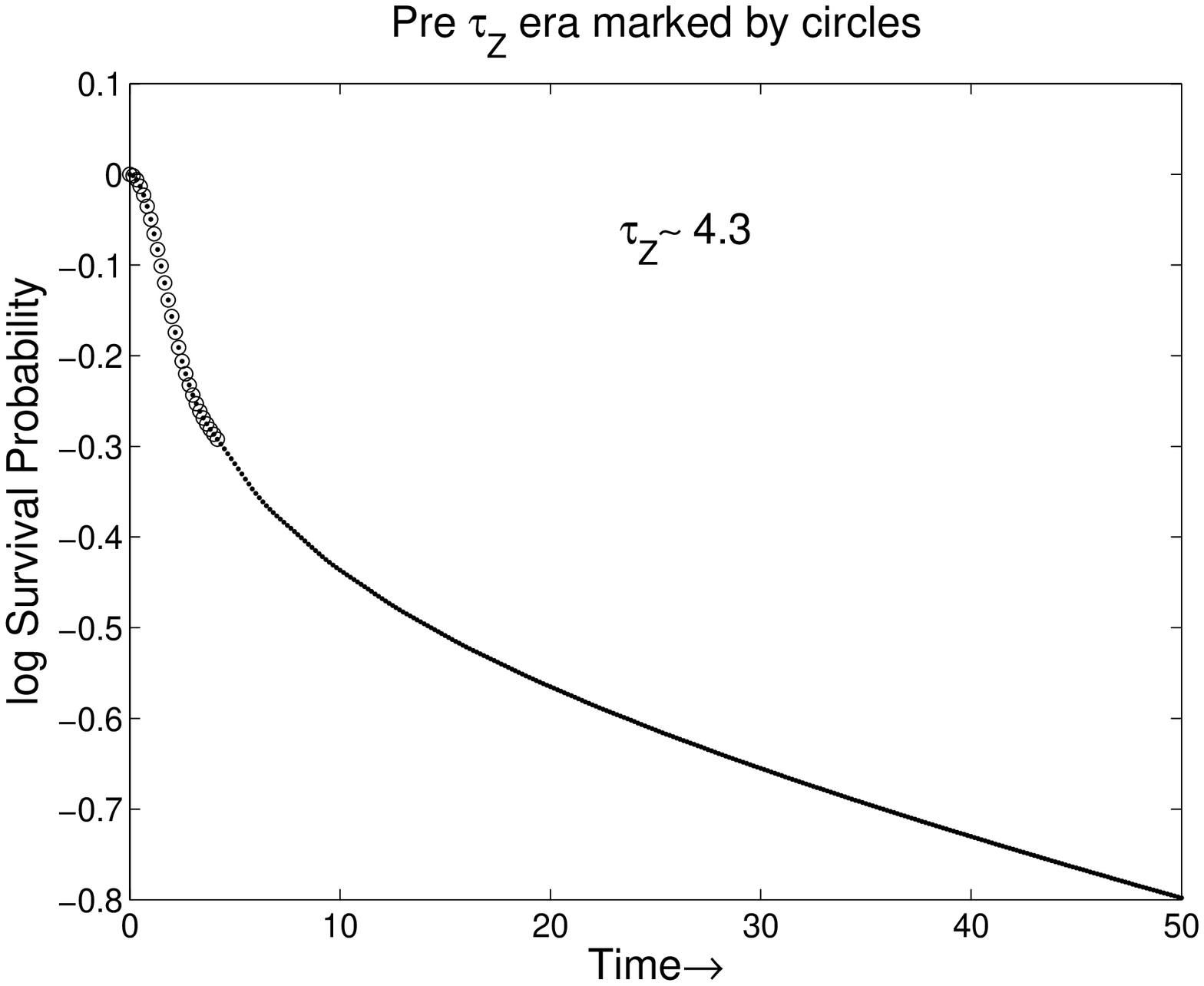}
\includegraphics[height=.25\textheight,width=.35\textwidth]{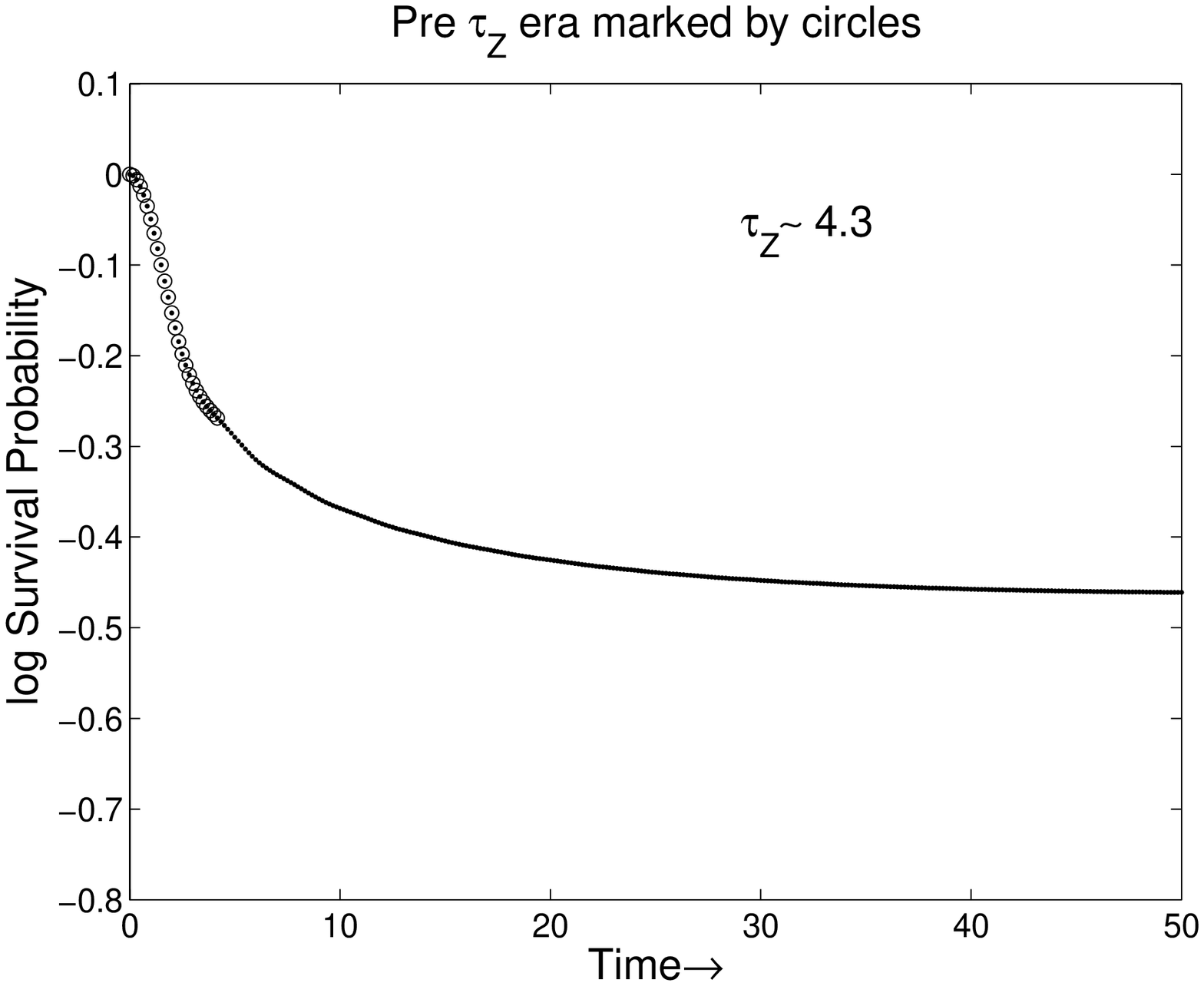}}
\caption{A short-time close-up of the decays in Fig.\ \ref{fig:plasmoncomparison}. For very short times one can see parabolic behavior (just barely discernable) and for longer times there is an anti-Zeno effect. All these peculiarities are phased out by about $\tauZ$, the Zeno time, defined in the text.}
\label{fig:plasmoncomparisonshorttime}\labeld{fig:plasmoncomparisonshorttime}
\end{figure}

\textbf{Remark}:~ Although the slight shift in $h$ above creates markedly different long-term behavior, for short time the quantum Zeno effect for the two parameter sets is similar. In Fig.\ \ref{fig:plasmoncomparisonshorttime} we show in greater detail the ``Zeno era,'' $t\leq \tauZ$, with $\tauZ$ what I have called the Zeno time \cite{jumppassage, jumpduration}, $\tauZ\equiv \hbar/ \sqrt{ \langle0|H^2|0\rangle - \langle0|H|0\rangle^2}$.

\textbf{Remark}:~ Formally the structure of the plasmon Hamiltonian and the structure of our $H_0$ and $V_I$ [Eqs.\ (\ref{eq:H01}/\ref{eq:H02}) and Eqs.\ (\ref{eq:V1}/\ref{eq:V2})] are similar. The matrix for finding the classical local modes has the form \Eqref{eq:decaymodel} if one first diagonalizes the ``chain'' as we do in Sec.\ \ref{sec:pathintegral}. The adjustable parameter sitting in the 1-1 position of the Hamiltonian of \Eqref{eq:decaymodel} is then $\omega_1^2$.

\begin{figure}
\includegraphics[height=.25\textheight,width=.35\textwidth]{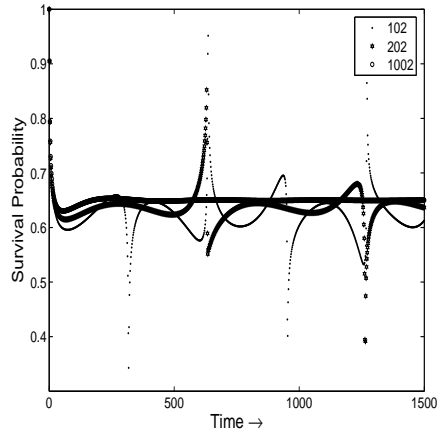}
\caption{Logarithm of survival probability with a plasmon for different matrix sizes. As is evident, the asymptotic probability is well defined, with the only affect of matrix dimension being the time at which finite-size recurrence begins (the quantum analogue of the Poincar\'e recurrence associated with finite-dimensional quasiperiodicity).}
\label{fig:NdepPlasmon}\labeld{fig:NdepPlasmon}
\end{figure}

Next we check that the decay inhibition due to the plasmon is independent of the matrix size. This is shown in Fig.\ \ref{fig:NdepPlasmon}, where the only change with increasing $N$ is the deferral of the Poincar\'e recurrence.

%%%%%%%%%%%%%%%%%%%%%%%%%%%%%%%%%%%%%%%%%%%%%%%%%%%%%%%%%
\subsection{Quantum time evolution of the breather\label{sec:quantumtimebreather}}
%%%%%%%%%%%%%%%%%%%%%%%%%%%%%%%%%%%%%%%%%%%%%%%%%%%%%

\begin{figure}
\centerline{\includegraphics[height=.25\textheight,width=.4\textwidth]{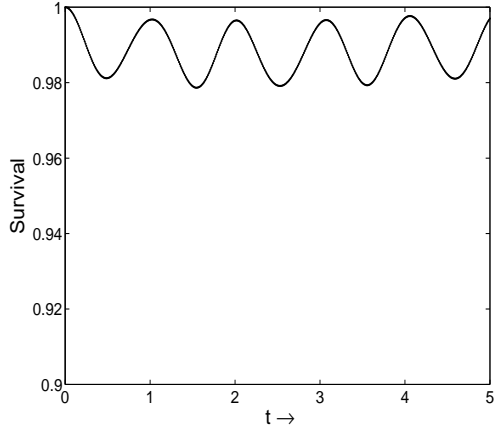}}%
\caption{Survival as a function of time; the initial state is the singly excited local mode.}
\label{fig:decay}
\end{figure}

We turn to the time dependence under quantum evolution of the breather state. Specifically, we study $|\langle\psi_0| \exp(-iHt/\hbar) |\psi_0\rangle|^2$ for $\psi_0$ a local mode phonon (in particular, $|0,\dots,1\rangle$) and $H$ the full Hamiltonian. This is shown in Fig.\ \ref{fig:decay} \cite{note:notthelargest}. As we saw in the examples of Sec.\ \ref{sec:generaldecay}, where a plasmon has formed, initial decay is followed by stabilization bounded away from zero. In the present situation the value at which it stabilizes is much larger than the $\sim$0.6 (cf.\ Fig.\ \ref{fig:NdepPlasmon}) of those examples, which reflects the much larger amplitude of the local mode eigenstate in the true eigenstate of the Hamiltonian (cf.\ Table \ref{tab:state}). Although the coupling pattern for the two cases is not the same (we will turn to this in a bit), the basic idea is the same: the initial state has order unity overlap with a true eigenfunction, and although it has coupling to the continuum, that coupling does not lead to exponential decay.

%%%%%%%%%%%%%%%%%%%%%%%%%%%%%%%%%%%%%%%%%%%%%%%%%%%%%%%%
\subsubsection{\label{sec:offdiaglocx}Off-diagonal matrix elements of the singly-excited local mode}
%%%%%%%%%%%%%%%%%%%%%%%%%%%%%%%%%%%%%%%%%%%%%%%%%%%%

Since for the plasmon mode the emphasis is on the threshold structure of the coupling, it is of interest to explore those properties for the breather. This is also useful for assessing properties of the eigenstate when, because of the size of the system, numerical diagonalization is out of reach.

\begin{figure}
\includegraphics[height=.25\textheight,width=.33\textwidth]{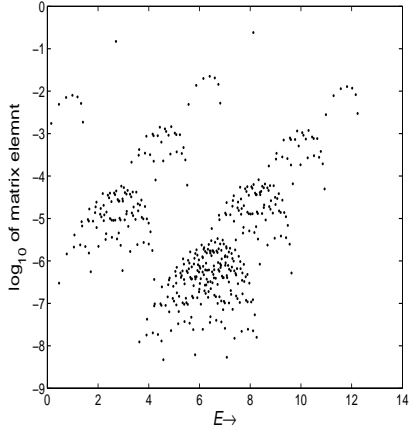}
\caption{Base-10 logarithm of $|\langle0,0,\dots,0,1|x_0^4|n_1,n_2,\dots\rangle|$ for a ring of 14 oscillators. States, $|n_1,n_2,\dots\rangle$, are ordered by increasing energy.}
\label{fig:matrixelements}\labeld{fig:matrixelements}
\end{figure}

States of the local mode are both our initial states for decay and our unperturbed states for the numerical diagonalization. One of these is $\local\equiv|0,0,0,\dots,1\rangle$ where, as in Sec.\ \ref{sec:diagonalization}, the ``1'' refers to the eigenvalue of the local mode number operator and the zeros to other phonon levels. The coupling, the analogue of ``$C$'' of \Eqref{eq:decaymodel}, is
\be
g(|n_1,n_2,\dots\rangle)\equiv\localdag x_0^4|n_1,n_2,\dots\rangle \,.
\label{eq:matrixelements}
\ee
\labeld{eq:matrixelements}
In Fig.\ \ref{fig:matrixelements} is a logarithmic plot of $g$ as a function of energy. With states ordered by energy [as in \Eqref{eq:decaymodel}] this clearly does not provide a smooth function. However, the principal issue is not a precise resemblance to the plasmon paradigm. The plasmon was discussed not because of its specific coupling pattern, but rather as an example of how, although the initial state \emph{does} couple to a continuum, the initial state still has finite (i.e., not going to zero with system size) overlap with a true eigenstate of the system. The pattern seen in Fig.\ \ref{fig:matrixelements}, shown for 14 oscillators, is the same as one sees for 4, 6, 8 or any number that we have been able to study. As such, the substantial overlap of the true eigenstate (induced by the nonlinear $\lambda$) with the local mode (induced by $\omega_1$) should continue as system size grows, since it these matrix elements that determine the overlap.

\textbf{Remark}:~ Note that stability may not persist in all dimensions; certainly threshold features of the density of states and spectrum are affected by dimension, and the usual intuitions regarding Fermi's Golden rule may again hold sway. In \cite{confine} we made the point that the symmetry breaking of the Jahn-Teller effect makes this a one-dimensional problem, significantly enhancing the possibility of classical breathers. The same may be true quantum mechanically. It may even be that this plays a role in the temperature-dependent decay of the breather (through an effective increase in dimension), as evidenced by the high-temperature disappearance of anomalous decay in doped alkali halides \cite{theoryandfit3}. For numerical diagonalization, increase in dimension is difficult computationally. However, for the path integral there should be little problem, and indeed 3-dimensional phonon coupling is used in many significant problems (see for example \cite{weiss}).

%%%%%%%%%%%%%%%%%%%%%%%%%%%%%%%%%%%%%%%%%%%%%%%%%%%%%%%%%%
\section{Conclusions\label{sec:conclusions}}
%%%%%%%%%%%%%%%%%%%%%%%%%%%%%%%%%%%%%%%%%%%%%%%%%%%%%%%%%

Both numerical diagonalization and path integration automatically deal with infinite-lifetime eigenstates of the Hamiltonian. The principal issue is therefore to show localization for the system we study. That system is not the full periodic system, but rather a version that neglects quantum-tunneling delocalization [an order $\exp(-[\hbox{positive constant}]/\hbar)$ effect], but otherwise differs little from the full translationally invariant system.

Localization is demonstrated in two ways. For the path integral we show that increasing value of the quartic coupling constant (``$\lambda$'') decouples the rapidly oscillating atom from distant atoms, in the same way the this phenomenon occurs for a quartic localized mode. For the method of numerical diagonalization we show that the breather state is, up to tiny corrections, entirely constructed of states of an appropriate quadratic local mode. We remark that there is no principle saying that such tiny corrections represent either delocalization or decay. For example, the harmonic oscillator state, $\psi(x)\sim\exp(-m\omega x^2/2)$ is certainly a localized object, even though it is nonzero for all $x$. As to decay, we devote an entire section to disabusing anyone of the notion that decay is a mathematical imperative, although what lies behind the non-decay is merely the fact that under the full Hamiltonian the overlap of a state $\psi_0$ with itself does not go to zero if $\psi_0$ has finite (bounded away from zero) overlap with an eigenstate of the full Hamiltonian.

It follows that the quantized breather is stable, except against spontaneous tunneling, a process that is in general far slower than the decay rates that have been ascribed to the breather.

%%%%%%%%%%%%%%%%%%%%%%%%%%%%%%%%%%%%%%%%%%%%%%%%%%%%%%%%%%%%%%
\acknowledgments
We thank Bernard Gaveau for useful discussions. This work was supported by NSF grant PHY 00 99471 and Czech grants ME 587 and GA AVCR A1010210. 
%%%%%%%%%%%%%%%%%%%%%%%%%%%%%%%%%%%%%%%%%%%%%%%%%%%%%%%%%%%%%%%%%%


\begin{thebibliography}{41}
\bibitem[{not({\natexlab{c}})}]{note:qxmcorrelation}
\bibinfo{note}{Of course what we call the $\langle qx_m\rangle$-correlation is
  really the $\langle q(x_m+x_{N+1-m})\rangle$-correlation (or $\langle
  q(x_m+x_{-m})\rangle$). Note too that there is no mathematical problem in
  exchanging the order of the operations in \Eqref{eq:correlationdefiniition}.}
\bibitem[{not({\natexlab{e}})}]{note:notthelargest}
\bibinfo{note}{This figure does not reflect our calculations of largest
  dimension, but uses a more severe cutoff, one that allows matrix
  exponentiation.}
\end{thebibliography}
\end{document}